\documentclass[twocolumn]{aastex631}
\usepackage{cleveref}
\usepackage{booktabs}
\usepackage{longtable}
\usepackage{array}

\makeatletter
\usepackage{etoolbox}
\patchcmd\H@refstepcounter{\protected@edef}{\protected@xdef}{}{}
\makeatother

\newcolumntype{i}{>{\scriptsize}r}

\shorttitle{A Distinct Population of Jetted-\acp{agn}}
\shortauthors{Kiehlmann et al.}

\graphicspath{{./}{figs/}}

\crefname{equation}{Eq.}{Eqs.}
\Crefname{equation}{Equation}{Equations}
\crefname{figure}{Fig.}{Figs.}
\Crefname{figure}{Figure}{Figures}
\crefname{table}{Table}{Tables}
\Crefname{table}{Table}{Tables}
\crefname{section}{Section}{Sections}
\Crefname{section}{Section}{Sections}

\usepackage[nolist,nohyperlinks]{acronym}
\begin{acronym}
 \acro{ads}[ADS]{Astrophysics Data System}
 \acro{agn}[AGN]{Active Galactic Nucleus}
 \acroplural{agn}[AGN]{Active Galactic Nuclei}
 \acro{bologna}[Bologna]{Bologna Complete Sample of Nearby Radio Sources}
 \acro{cd}[CD]{Compact Double}
 \acro{cjf}[CJF]{Caltech Jodrell Bank flat-spectrum}
 \acro{class}[CLASS]{Cosmic Lens All-Sky Survey}
 \acro{cso}[CSO]{Compact Symmetric Object}
 \acro{css}[CSS]{Compact Steep Spectrum}
 \acro{ddrg}[DDRG]{Double-Double Radio Galaxy}
 \acroplural{ddrg}[DDRGs]{Double-Double Radio Galaxies}
 \acro{emerlin}[eMERLIN]{Enhanced Multi Element Remotely Linked Interferometer Network}
 \acro{fsrq}[FSRQ]{Flat Spectrum Radio Quasar}
 \acro{fwhm}[FWHM]{Full Width at Half Maximum}
 \acro{ps}[PS]{Peaked Spectrum}
 \acro{hfp}[HFP]{High-Frequency Peaker}
 \acro{ks}[KS]{Kolmogorov-Smirnov}
 \acro{mso}[MSO]{Medium Symmetric Object}
 \acro{ned}[NED]{NASA/IPAC Extragalactic Database}
 \acro{ovro}[OVRO]{Owens Valley Radio Observatory}
 \acro{pr}[PR]{Pearson Readhead}
 \acro{ps}[PS]{Peaked-Spectrum}
 \acro{prcj1}[PR+CJ1]{Pearson Readhead and First Caltech Jodrell Bank}
 \acro{rfc}[RFC]{Radio Fundamental Catalog}
 \acro{sfr}[SFR]{Star Formation Rate}
 \acro{smbh}[SMBH]{Supermassive Black Hole}
 \acro{vips}[VIPS]{VLBA Imaging and Polarimetry Survey}
 \acro{vla}[VLA]{Very Large Array}
 \acro{vlba}[VLBA]{Very Long Baseline Array}
 \acro{vlbi}[VLBI]{Very Long Baseline Interferometry}
\end{acronym}

\begin{document}

\title{Compact Symmetric Objects - II \\  Confirmation of  a Distinct Population of  High-Luminosity  Jetted Active Galaxies}

\correspondingauthor{Anthony Readhead}
\email{acr@caltech.edu}

\author{S. Kiehlmann}
\affiliation{Institute of Astrophysics, Foundation for Research and Technology-Hellas, GR-70013 Heraklion, Greece}
\author{A. C. S Readhead}
\affiliation{Owens Valley Radio Observatory, California Institute of Technology, Pasadena, CA 91125, USA}
\author{S. O'Neill}
\affiliation{Owens Valley Radio Observatory, California Institute of Technology, Pasadena, CA 91125, USA}
\author{P. N. Wilkinson}
\affiliation{Jodrell Bank Centre for Astrophysics, University of Manchester, Oxford Road, Manchester M13 9PL, UK} 
\author{M. L. Lister}
\affiliation{Department of Physics and Astronomy, Purdue University, 525 Northwestern Avenue, West Lafayette, IN 47907, USA}
\author{I. Liodakis}
\affiliation{Finnish Center for Astronomy with ESO, University of Turku, Vesilinnantie 5, FI-20014, Finland}
\affiliation{Department of Physics, Univ. of Crete, GR-70013 Heraklion, Greece}
\author{S. Bruzewski}
\affiliation{Department of Physics and Astronomy, University of New Mexico, Albuquerque, NM 87131, USA}
\author{V. Pavlidou} 
\affiliation{Institute of Astrophysics, Foundation for Research and Technology-Hellas, GR-70013 Heraklion, Greece}
\affiliation{Department of Physics and Institute of Theoretical and Computational Physics, University of Crete, 70013 Heraklion, Greece}
 \author{T. J. Pearson}
\affiliation{Owens Valley Radio Observatory, California Institute of Technology, Pasadena, CA 91125, USA}
\author{E. Sheldahl}
\affiliation{Department of Physics and Astronomy, University of New Mexico, Albuquerque, NM 87131, USA}
\author{A. Siemiginowska}
\affiliation{Center for Astrophysics|Harvard and Smithsonian, 60 Garden St., Cambridge, MA 02138, USA}
\author{K. Tassis} 
\affiliation{Institute of Astrophysics, Foundation for Research and Technology-Hellas, GR-70013 Heraklion, Greece}
\affiliation{Department of Physics and Institute of Theoretical and Computational Physics, University of Crete, 70013 Heraklion, Greece}
\author{G. B. Taylor}
\affiliation{Department of Physics and Astronomy, University of New Mexico, Albuquerque, NM 87131, USA}

\begin{abstract}
 \acfp{cso}  are compact ($<1$ kpc), jetted \acp{agn}, whose jet axes are not aligned close to the line of sight, and whose observed emission is not predominantly relativistically boosted towards us.  Two classes of CSOs have previously been identified: approximately one fifth are edge-dimmed and designated as CSO 1s, while the rest  are edge brightened and designated as CSO 2s. This paper focuses almost exclusively on CSO 2s. Using complete samples of  CSO 2s we present three independent lines of evidence, based on their relative numbers, redshift distributions, and size distributions,  which show conclusively that the vast majority ($> 99\%$) of CSO 2s do not evolve into larger-scale radio sources. These CSO 2s  belong to a distinct population of  jetted-\acp{agn}, which should be characterized as ``short-lived''  compared to the classes of larger jetted-AGN, as opposed to ``young''.  We show that there is a sharp upper cutoff in the  CSO 2 size distribution at $\approx 500$ pc. The distinct differences between   most CSO 2s and other jetted-\ac{agn}   provides a crucial new time domain window on the formation and evolution of relativistic jets in \acp{agn} and the supermassive black holes that drive them.

\end{abstract}


\section{Introduction}
\label{sec:intro}

The first indication of relativistic motion in the jets of active galaxies was the asymmetric large-scale jet in M87 discovered by \citet{1918PLicO..13....9C}. The next was arguably the discovery of rapid flux density variations in blazars \citep{1965Sci...148.1458D,1965AJ.....70..672D}, which were quickly shown by \citet{1966Natur.211..468R,1967MNRAS.135..345R} to be due to relativistic motion of the emission regions towards the observer. In spite of this development, 
observations of the synchrotron self-absorption cutoff frequencies of radio sources with flat spectra led to the hypothesis that an ``inverse Compton catastrophe'' imposes an upper limit of $\sim 10^{12}\,\mathrm{K}$ on the brightness temperatures of compact radio sources \citep{1969ApJ...155L..71K}. This appeared, at first, to be supported by \ac{vlbi} observations, but in these calculations the possibility of relativistic bulk motion towards the observer \citep{1966Natur.211..468R,1967MNRAS.135..345R} was not taken into account. 

 The first phase-coherent astronomical image ever obtained in any energy band, including the optical band, having a resolution significantly less than 
 one arc second was produced in the first ``hybrid map'',  which showed an asymmetric  one-sided radio jet \citep{1977Natur.269..764W}. Such core-jet structures were soon shown to predominate in compact radio sources \citep{1978Natur.276..768R,1980IAUS...92..165R}, making it clear that relativistic beaming determines the apparent morphology  and the observed brightness temperatures of most compact radio sources at cm wavelengths. Nevertheless, the  inverse Compton catastrophe hypothesis continued to propagate, but,  as shown by \citet{1994ApJ...426...51R}, when relativistic beaming is taken into account, the brightness temperatures drop to $\sim 10^{11}\,\mathrm{K}$, and are consistent with equipartition between the magnetic field and particle energy densities in the emission regions.

 It should therefore be clear that relativistic beaming greatly complicates the physical analysis of the observed radio emission of compact radio sources. 
In order to overcome such complications, which introduce large uncertainties in the physical properties, such as the  magnetic field strength,  the particle energy densities, the pressures, and the total energies of the emission regions, \citet{1994ApJ...432L..87W}, hereafter W94, introduced the \acf{cso} classification of compact radio sources.
Due to the morphological symmetry of the emission on either side of the nucleus, these objects are clearly not exhibiting strongly beamed emission towards the observer.

Unfortunately, a number of jetted-\acp{agn} have been misidentified  as CSOs or \ac{cso} candidates in the literature, and many jetted-\acp{agn} whose axes are close to the line of sight, and whose observed emission is strongly beamed towards us, have crept into this class. This paper is the second of three  on the morphological radio properties of CSOs in which we explore CSO phenomenology uncontaminated by objects that have been mis-identified as CSOs.  In the first paper (Paper~1: Kiehlmann et al. in press) we added two new criteria, based on variability and speed, to the \ac{cso} selection criteria and undertook a detailed survey of the literature, which enabled us to identify 79~bona fide CSOs.   From the 79 bona fide CSOs we determined the numbers  in three complete  samples\footnote{ A ``complete sample'' is defined to be a sample that includes all objects down to a given flux density limit over a given area of sky \citep{1968MNRAS.139..515P,1968ApJ...151..393S,1970MNRAS.151...45L}} from which,  in this paper (Paper~2), we show that $\gtrsim 99\%$  of CSO 2s form a class of jetted-\acp{agn} that is both distinct from other jetted-\acp{agn} and  exhibits a sharp cutoff in size at $\approx 500$ pc, and a corresponding cutoff in age at $\approx 5000$ yr, so that only fewer than  $ 1\%$ of CSO 2s might possibly go on to form the larger classes of jetted-AGN, such as Fanaroff and Riley Type I (FRI) and Type II (FR II) objects \citep{1974MNRAS.167P..31F}.  In the third paper (Paper~3: Readhead et al. in press) we discuss the evolution of CSO 2s and show that while they are nearly all  ``short-lived'' compared to the classes of larger jetted-AGN, only a minority of them are ``young''.   Note that FR~I and FR~II objects have sizes in the $\sim$20 \,kpc -- several Mpc range, and therefore clearly have ages much longer than the vast majority of CSO 2s. 
We should avoid the implicit assumptions involved in calling  all  CSO 2s ``young'', which obscure the true nature and  importance of this class of jetted-\acp{agn}. It is critically important, therefore, to recognize the distinction between the terms  ``young'' and ``short-lived'', which otherwise obfuscate the phenomenology of the \ac{cso} class.

  In an important development in the study of CSOs, \citet{2016MNRAS.459..820T} showed that there are two major morphological classes of CSOs: edge-dimmed objects, which we designate as CSO 1s, and edge-brightened objects, which we  designated as CSO 2s. Paper 1 confirms their finding and this paper deals almost exclusively with CSO 2s.

As discussed in detail in Paper~1, in CSOs, two emission regions are seen straddling the center of activity, making it clear that these cannot be strongly relativistically boosted, otherwise the object would be seen as a one-sided asymmetric ``core-jet'' object as is the case in the vast majority of compact radio sources observed at cm wavelengths \citep{1998AJ....115.1295K,2019ApJ...874...43L}. 

Individual CSO 2s undergo appreciable evolutionary  structural changes on timescales of years that can therefore be studied without the complications of relativistic beaming.  The bulk flows along their jets and their speeds of advance into the interstellar medium can be measured directly. We argue that CSO 2s provide a uniquely accessible  time domain laboratory for the study of relativistic jets \citep{2019ARAandA..57..467B} and the SMBH central engines that drive them, because they are short-lived compared to the classes of larger jetted-AGN, rather than young, and hence pass through all stages of their lives as CSO 2s, which are  therefore available for detailed study in all phases of their lives.  It is important to distinguish between CSOs that have small sizes because they are ``short-lived'' compared to larger classes of jetted-AGN, and CSOs that have been stalled by the interstellar medium of their host galaxies and therefore stopped growing in size. We propose the hypothesis that such stalled CSOs are likely to be edge-dimmed and hence fall into the CSO 1 class.  We also note in passing that this could be of great importance to feedback.  As we show in Paper 3, the most luminous CSO 2s that are the subject of this study have not been stalled -- their hot spots are separating on average at $\sim 0.4 c$, and their maximum lifetimes are $\sim 5000$ yr.  For the purposes of this study, although stalled CSO 2s are of great potential interest, we do not consider them further in these three papers.  A minority of the less luminous CSO 2s in our study might possibly be stalled systems and should also be considered in that light. But this is beyond the scope of the present study.

By the early 1990s, three bona fide CSO 2s had been definitively identified in the complete sample of 65 radio sources studied by \citet{1988ApJ...328..114P}. Despite the small size of the \ac{cso} sample, and entirely because it was part of a complete sample, this  sample of only three CSO 2s was enough to enable a number of the most critical questions about CSO 2s to be addressed by \citet{1994cers.conf...17R}, hereafter R94, including their relationship to the larger jetted-AGN, their lifetimes, and their energy requirements. R94 concluded that CSO 2s form a distinct population of compact jetted-\acp{agn}, and that there must be a physical reason for this which provides a unique window on the central engines that drive \acp{agn}. R94 also  suggested that CSO 2s might be the result of the capture of a single star by a SMBH in an otherwise quiescent
elliptical galaxy nucleus.  This possibility was also suggested more recently by \citet{2012ApJ...760...77A}.

All of these properties of 
 CSO 2s were discussed in more detail, and confirmed,  in \citet{1996ApJ...460..612R}, hereafter R96. Nevertheless, in spite of their distinction, CSO 2s have attracted comparatively little attention among jetted-\acp{agn} enthusiasts. We explore the characteristics of CSO 2s in considerably more detail in this paper and in Paper~3.

   The CSOs are a subset of AGN, but by studying a restricted well-defined sample of CSOs we aim to understand them in depth and gain new insights into the physics and formation of jetted-AGN. Although  it is not a primary goal of these papers, we discuss the relationship of CSOs to other classes of AGN where appropriate in this paper and in Paper 3.  To place CSOs in the broader context of compact radio sources associated with AGN, the reader is referred to the comprehensive review of \citet{2021AARv..29....3O}, hereafter OS21.

\begin{deluxetable}{llll}
\tablecaption{The CSO Samples}
\tablehead{ & CSO~1 & CSO~2 & All}
\startdata
    With spectroscopic redshift    & 11 & 43 (17) & 54 \\
    Without spectroscopic redshift & 5  & 20 (2)  & 25 \\
    All                            & 16 & 63 (19) & 79 \\
\enddata
\tablecomments{This Table shows the numbers of bona fide CSOs of classes~1 and~2 identified in Paper 3, with and without spectroscopic redshifts. Numbers in parentheses indicate bona fide CSOs in the PR+CJ1+PW complete samples. In this paper we deal almost exclusively with the 17~CSOs in these complete samples that have spectroscopic redshifts.}
\label{tab:classes}
\end{deluxetable} 

Throughout this paper we adopt the convention $S_\nu \propto
\nu^{\alpha}$ for spectral index $\alpha$, and use the cosmological
parameters $\Omega_\mathrm{m} = 0.27$, $\Omega_\Lambda = 0.73$ and $H_0 = 71 \;
\mathrm{km\; s^{-1} \;\,Mpc^{-1}}$ \citep{Komatsu09}. We do this for consistency with our other papers. None of the conclusions would be changed were we to adopt the best model of the Planck Collaboration  \citep{2020AandA...641A...6P}.

\begin{figure}[!t]
 \centering
 \includegraphics[width=1.0\linewidth]{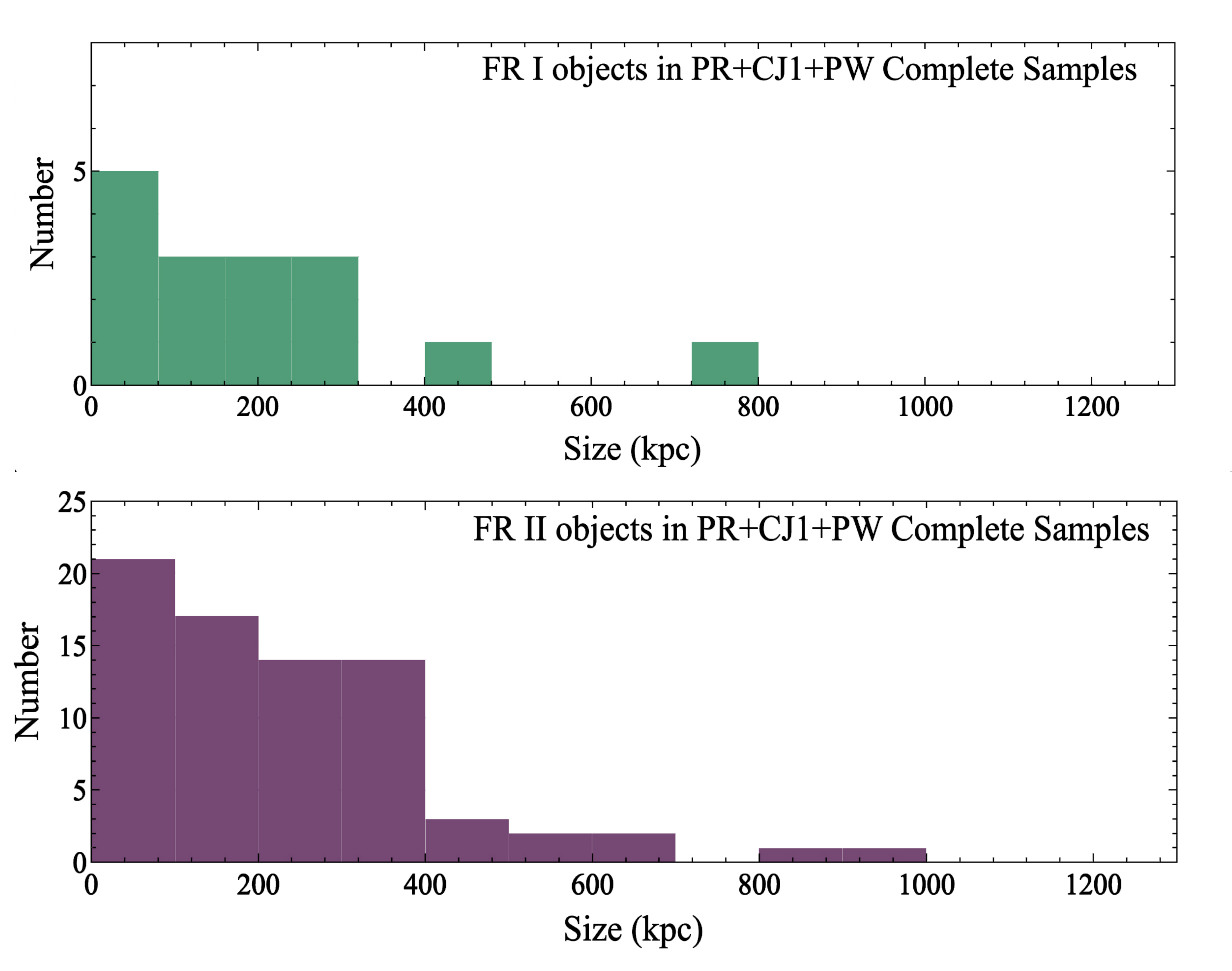}
 \caption{The  distributions of the largest projected sizes of FR~I objects (top panel) and FR~II objects (bottom panel) in the PR+CJ1+PW complete samples. The FR I sizes have been determined from our own angular size measurements. The FR II sizes are based on the largest angular size measurements of \citet{1993ApJ...413..453N} for all but six sources not included in their sample, for which we measured the angular sizes ourselves. 
 There is one FR~II object (3C 236) of size 4.3\,Mpc that is not included in the FR~II plot.  }
 \label{plt:histogramlargescale}
\end{figure}

\section{Complete Samples of Jetted-AGN}\label{sec:complete}

  The disposition of the CSOs we consider in this study, amongst the CSO 1 and CSO 2 classes, the CSOs with spectroscopic redshifts and those without, and the CSOs in complete samples is shown in Table \ref{tab:classes}.  The classification of CSO 1s and CSO 2s that we use here is discussed in more detail in Paper 3.

 We use only complete samples for statistical tests in this paper. Other methods for making statistical tests, which are not based on complete samples,  must introduce some assumptions regarding the population under study, and we wish to avoid making such assumptions. We use three complete samples extensively (see Table \ref{tab:csonumbers}): (1) The 5 GHz Pearson--Readhead (PR) complete sample \citep{1988ApJ...328..114P} based on the MPIfR/NRAO~S4 and S5~surveys \citep{1978AJ.....83..451P,1981AJ.....86..854K}; (2) The first Caltech--Jodrell (CJ1) 5 GHz complete sample \citep{1995ApJS...98....1P,1995ApJS...99..297X}l;
(3) The Peacock--Wall (PW) 2.7 GHz complete sample \citep{1981MNRAS.194..331P,1985MNRAS.216..173W}.\footnote{The original PW sample \citet{1981MNRAS.194..331P} contained 168 sources; to these three (DA~240, 0945+73 = 4C~73.08, and NGC~6251) were added by \citet{1985MNRAS.216..173W}.}

There are 282~objects in the union of these three complete samples and these are listed in \cref{tab:samples} in the Appendix. In our analysis in this paper, we exclude M82 (3C~231) which is in  the PR and PW 
 samples, but is a starburst galaxy and not a jetted AGN, leaving  281 sources.  In Paper 1 we listed the number of CSOs in the three complete samples. The number of CSO 2s in each of these three complete samples is given in \cref{tab:csonumbers}. The determination of a uniform set of measurements of the largest angular size of the 79 bona fide CSOs is described in Paper~1.

All the  sources in the PW sample were mapped using the Cambridge 5 km Telescope by \citet{1981MNRAS.194..331P}, who also classified the large scale structures in  the PW sample according to the following types:~(i) FR~I and FR~II, and an intermediate FR~type (FR?); (ii) objects unresolved on the 5 km Telescope (U); (iii)~\ac{css} objects having $\alpha \le -0.5$ between 2.7\,GHz and 5\,GHz; and (iv)~double objects with the optical identification coincident with one of the two radio components.  These types are listed in column 7 of \cref{tab:samples} in the Appendix.

Discussions of a size cutoff in CSO 2s are not new \citep{1998MNRAS.299.1159A,2006MNRAS.368.1411A,2009AN....330..190A}, and early lobe-speed measurements showed that the hotspots of CSO 2s are rapidly separating  \citep{1998AandA...337...69O,1999NewAR..43..669O, 2002evn..conf..139P}. It was clear, therefore, as pointed out in R94 and R96, that CSO 2s must be short-lived, since otherwise there would be far more of their longer-lived, larger counterparts. This means that CSO 2s {\it must} exhibit a size cutoff. As shown in this paper,  we have now determined that a sharp cutoff occurs at $\approx 500$~pc. The evolution of  the vast majority of CSO 2s  from ``early-life'' through ``mid-life'' to ``late-life'' is discussed in detail in Paper~3,  where we also discuss the fact that a small fraction  ($<1\%$) of CSO 2s almost certainly go on to form the larger classes of jetted-AGN, including MSOs, FR Is and FR IIs.

 Since the CSOs in the PR+CJ1+PW complete samples
 are all CSO 2s, the findings of this paper apply, (i)  only to CSO 2s, and (ii) only to the high-luminosity end of the CSO 2 luminosity function. Much deeper complete sample surveys, in which we are engaged, are needed to expand our knowledge into the low-luminosity regime of CSO 2s.  It should therefore be born in mind that our sample, comprising only 17 objects, is small, so that a degree of caution must be exercised in interpreting the results.  For this reason we present all of the relevant statistics and p-values so that readers may judge for themselves the significance of the results.

There are precedents in astronomy for drawing powerful conclusions based on small numbers. For example, \citet{1929PNAS...15..168H,1929CoMtW...3...23H}, based his discovery of the expansion of the universe on measurements of just 22 galaxies. \citet{1941PASP...53..224M} had just 14 supernovae for his classification of Type I and Type II  supernovae, with 9 and 5 objects, respectively. Closer to the approach of this paper, there is also a powerful precedent for using well-defined statistical tests based on complete samples in the paper by \citet{1968ApJ...151..393S}, who used a complete sample of just 33 quasars to demonstrate convincingly that quasars are not evenly distributed in space, but show strong cosmological evolution.

\begin{deluxetable*}{c@{\hskip 8mm}ccccccccc}
\tablecaption{The Numbers of CSS, FR~I, FR~II, and CSO objects in the Complete Samples. }
\tablehead{Complete& Flux Density&CSS&FR~I&FR~II&Total&CSO2&CSO2/FR~II&CSO2/Total\\
Sample&limits&Number&Number&Number&Number&Number&Percentage&Percentage}
\startdata
PR&$S_5\geq 1.3$\,Jy&6 &3&16&$64^\dag$&6&37.5$\pm$18.0 \%&9.4$\pm$4.0 \%\\
CJ1&1.3\,Jy $\geq S_5 \geq 0.7$\,Jy&23 &6&30&135&$6^\ddag$&20.0$\pm$8.9 \%&4.4$\pm$1.9 \%\\
PR$+$CJ1&$S_5\geq 0.7$\,Jy&29 &8&46&$199^\dag$&$12^\ddag$&26.1$\pm$8.5 \%&6.0$\pm$1.8 \%\\
PW&$S_{2.7}\geq 1.5$\,Jy&26 &15&65&$170^\dag$&13$^*$&20.0$\pm$6.1 \%&7.6$\pm$2.2 \%\\
PWS&$S_5\geq 1.3$\,Jy&7 &8&11&50&5&45.5$\pm$24.5 \%&10.0$\pm$4.7 \%\\
PR$+$CJ1$+$PW& - & 43 &16&76&$281^\dag$&$19^\ddag$ $^*$&25.0$\pm$6.4 \%&6.8$\pm$1.6 \%\\
\enddata
\tablecomments{All of the CSOs in the PR+CJ1+PW sample are CSO 2s. In addition, all of the PW CSOs at $\delta \geq 35^\circ$  are in the PD+CJ1 sample. $^\dag$ the numbers exclude 3C~231 (M82), a starburst galaxy, not an AGN. $^\ddag$ the numbers include the bona fide \ac{cso} J1335+5844, for which there is no published spectroscopic redshift. $^*$ the numbers include the bona fide \ac{cso} J1416+3444, for which there is no published spectroscopic redshift.  These numbers are taken from the list of the 282~sources in the three complete samples given in \cref{tab:samples} in the Appendix. PWS is the subsample of PW at $10^\circ<\delta<35^\circ$ (B1950) and with $S_{5\,\mathrm{GHz}}> 1.3$\,Jy.  As should be clear in view of the size of the samples, and assuming there is no dependence of  the CSO fraction on flux density,  the most reliable statistic is the final one combining the three full samples PR, CJ1, and PW. }
\label{tab:csonumbers}
\end{deluxetable*}

In this paper we present three independent sets of data and lines of argument based on complete samples, each of which shows that the vast majority of  CSO 2s form a distinct population of jetted-AGN. These lines of argument are based on (i) the numbers of CSO 2s in complete samples; (ii) the redshift distributions of these CSO 2s; and (iii) the size distribution of these CSO 2s.  Of these, the  results of first and third arguments are, in our view, compelling.  The results of the second argument (ii) are significant only at the p-value=$1.6 \times 10^{-2}$ ($2.1\sigma)$ level, and are, therefore, not compelling, but they are in the same sense as the other two arguments -- i.e. they strongly suggest that the CSO 2s are drawn, predominantly, from a distinct population of jetted-AGN.

\begin{deluxetable*}{c@{\hskip 8mm}cccccc}
\tablecaption{Redshifts and Sizes of the 17~CSO 2s with spectroscopic redshifts in the PR, CJ1 and PW Complete Samples}
\tablehead{IAU Name&Redshift&Size (pc)& PR&CJ1&PW\\}
\startdata
J0029+3456	&	0.517	&	180	&		&		&	Y	\\
J0111+3906	&	0.668	&	56	&	Y	&		&		\\
J0119+3210	&	0.0602	&	115	&		&		&	Y	\\
J0405+3803	&	0.05505	&	44	&		&	Y	&		\\
J0713+4349	&	0.518	&	217	&	Y	&		&	Y	\\
J1035+5628	&	0.46	&	221	&	Y	&		&	Y	\\
J1227+3635	&	1.975	&	499	&		&	Y	&	Y	\\
J1244+4048	&	0.8135	&	529	&		&	Y	&		\\
J1326+3154	&	0.37	&	345	&		&		&	Y	\\
J1347+1217	&	0.121	&	215	&		&		&	Y		\\
J1400+6210	&	0.431	&	378	&	Y	&		&	Y	\\
J1407+2827	&	0.077	&	16	&		&		&	Y		\\
J1609+2641	&	0.473	&	362	&		&		&	Y		\\
J1735+5049	&	0.835	&	61	&		&	Y	&			\\
J1944+5448	&	0.263	&	196	&		&	Y	&			\\
J2022+6136	&	0.227	&	104	&	Y	&		&	Y		\\
J2355+4950	&	0.237	&	337	&	Y	&		&	Y	\\
\enddata
\tablecomments{References for the redshifts and sizes are given in Paper~1.}
\label{tab:zandsize}
\end{deluxetable*}

The PW sample was selected at 2.7\,GHz, unlike the PR and CJ1 samples which were selected at 5\,GHz. However, we have 5\,GHz flux densities for all the PW sources \citep{1977IAUS...74...63P}.    Following a suggestion by John Peacock, in order to be able to combine results from these three complete samples without introducing any possible biases due to the different sample selection frequencies, we define a subset of the PW sample that is effectively complete at 5\,GHz. 
For this purpose we compare the GB6 \citep{1996ApJS..103..427G}, PR, and PW samples at 5\,GHz over their common sky area ($35^\circ \leq \delta \leq 75^\circ$, $|b| \geq 10^\circ$, B1950). These surveys were all made on different instruments at different times and since many of the sources are variable the samples change slightly with time. 
In this area of sky, the GB6 survey has 54, the PR survey  has 51, and the PW survey has 54 objects with  $S_{\rm 5\,GHz} \geq 1.3$\,Jy. 
It may safely be assumed, therefore, that the PW sample is effectively complete down to 1.3\,Jy at 5\,GHz. 
Of these we define a sub-sample, PWS, where ``S'' stands for ``Subsample'', consisting of the PW sources at declinations $\delta < 35^\circ$ (B1950) and having $S_{\rm5\,GHz} \geq  1.3$\,Jy, for use in our physical size distribution statistical tests in \S \ref{sec:size}.   We point out that all of the PW CSOs at $\delta \geq 35^\circ$ are in the PW+CJ1 sample.

\begin{figure*}[!t]
 \centering
 \includegraphics[width=1.0\linewidth]{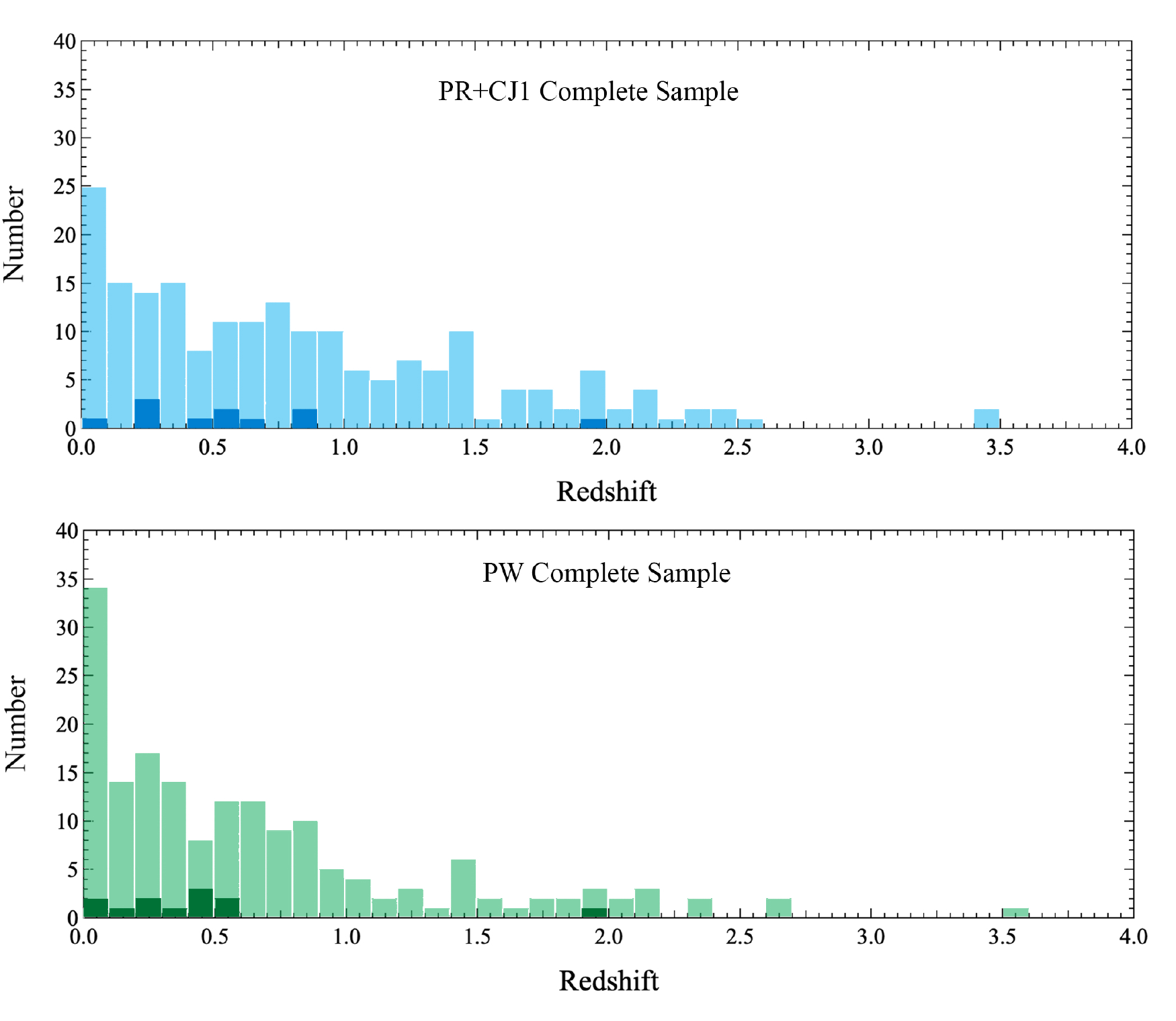}
 \caption{The redshift distributions for the PR+CJ1 complete sample (top panel) and the PW complete sample (bottom panel).  The light shaded  distributions show the complete samples. The dark shaded regions show the CSO 2s.  Note that these distributions are not stacked vertically, so the values on the ordinate represent the total numbers of sources and the numbers of CSO 2s in each sample. The cumulative distributions and KS statistics are shown in Figs.  \ref{plt:KSredshift} (a) and (b). }
 \label{plt:histogramredshift}
\end{figure*}

The great strength of the PR, CJ1, and PW samples is that {\it all} of the objects are well-studied and their radio properties on both large and small scales are known.   There is, therefore, no danger of unknown selection bias that could compromise the statistics. In \cref{tab:samples} in the Appendix we list all of the sources in the complete PR, CJ1 and PW samples and we provide references to these structure observations.   Clearly, the  references given in Table \ref{tab:samples} do not  include all of the papers that refer to the objects in these samples   -- in many cases we provide only a single reference to a paper that contains a good map of the object.


In addition to the above three complete samples, there is one other complete sample that is of prime importance to this study:
the GaLactic and Extragalactic All-Sky Murchison Widefield Array (GLEAM) survey \citep{2017ApJ...836..174C}, which covers the sky area $\delta<30^\circ$ (J2000), $|b|>10^\circ$ and defines a complete sample of 11,400~objects exhibiting flux densities greater than 1\,Jy in the 72\,MHz -- 700\,MHz range.

 In the following sections, using the complete samples,  we provide three lines of argument that the vast majority of  CSO 2s form a distinct population of jetted-AGN.

\begin{figure*}[!t]
 \centering
 \includegraphics[width=1.0\linewidth]{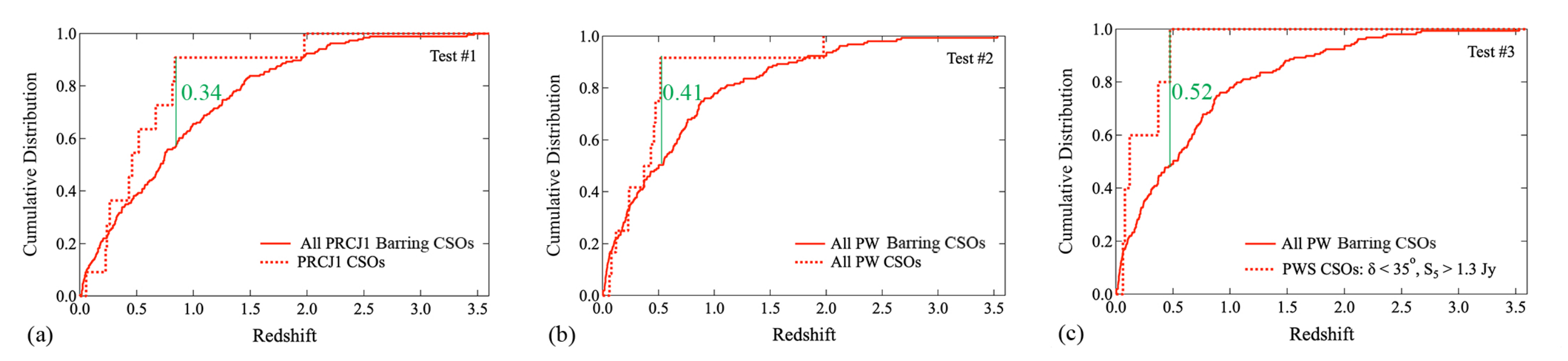}
 \caption{KS Tests on the redshift distributions of the bona fide CSO 2s in the PR+CJ1, PW, and PWS samples. (a), (b) and (c): comparison of the \ac{cso} cumulative redshift distributions {\it vs.} the non-CSO 2s in the  complete PR+CJ1, PW, and PWS samples, respectively.  The green bars indicate the maximum differences in the cumulative distributions, corresponding to the values of the KS statistic given by the numbers in green. The corresponding p-values are listed in \cref{tab:ksredshift}. }
 \label{plt:KSredshift}. 
\end{figure*}

\section{The Fractions of CSO 2\lowercase{s} in Complete Samples}\label{sec:statistics}

 It is important to note that, in addition to the identification of 79  ``bona fide'' CSOs in Paper~1, we also identified 167 ``class A'' CSO candidates, which are  objects showing clear double structure, but for which the maps are not of sufficient quality to confirm them as CSO 2s.  We have VLBA observations of these and are in process of analyzing them.  We also identified 1164  ``class B'' CSO candidates, most of which are far less likely to be CSO 2s, but which cannot yet definitively be ruled out.

 R96 gave a detailed discussion of the \ac{cso} fractions in the PR and CJ1 complete samples.  Here we update this discussion  and also incorporate the PW sample. 

While the PR \ac{cso} sample is complete, as can be seen in Paper~1, in CJ1 there are 5 class~A \ac{cso} candidates which might possibly be bona fide CSO 2s.  All of these candidates have sizes less than 500\,pc, and, were we to include these five sources in our analysis, the conclusions below would be strengthened.  We prefer to take the conservative route and not to include any CSO candidates in the bona fide sample until they have met the CSO criteria laid out in Paper~1.

As can be seen in Paper~1, there are also six class~B \ac{cso} candidates in the PR, CJ1, and PW samples.  These are much less likely to be bona fide CSO 2s, and all but one have sizes less than 500\,pc.  For these reasons, the conservative approach is again not to include any of these objects in the present analysis.

We see in \cref{tab:csonumbers} that the fraction of CSO 2s in complete 2.7--5\,GHz samples is $(6.8\pm 1.6) \%$. This would rise to $ (8.5 \pm 1.8)\% $ if all of the class~A CSO 2s candidates in the CJ1 sample are shown to be bona fide CSOs.

We take as a simple hypothesis to be used throughout this paper, that, between their appearance and disappearance, the separation speed, $v_{\rm sep}$, of the opposing hot spots in CSO 2s, when averaged over a sufficient interval of time, is constant, and that they continue at the same separation speed if they expand to form larger classes of sources, such as FR~IIs. Under this hypothesis, the number of objects in different size ranges scales  simply in proportion to the size ranges.

It is important to note that, for the purposes of our arguments regarding the fractions of CSO 2s with respect to classes of larger sources,  this hypothesis is highly conservative. We show in Paper~3 that the separation  speed of the  hot spots for CSO 2s  is $v_{\rm sep} =(0.36\pm0.04) $c, whereas, for example, \citet{1991ApJ...383..554C} argue convincingly that for the opposing  hot spots in Cygnus~A,   $  0.01 {\rm c } < v_{\rm sep} < 0.05 {\rm c}$. Note that this deduced separation speed for Cygnus~A is typical for FR IIs \citep{1995MNRAS.277..331S}.  Based on these values, the separation speeds in CSO 2s are approximately an order of magnitude greater than those in FR~IIs, which means that if CSO 2s do expand to form FR~IIs, they spend far less of their time in the 0\,kpc  to 1\,kpc size range than under the constant speed hypothesis, and so there should be even fewer of them relative to FR~IIs than under the constant speed hypothesis.

Under the constant speed hypothesis  we also assume  that the luminosity does not change enough for the source to drop out of the flux-limited sample. We discuss possible changes in luminosity later.

\begin{deluxetable*}{c@{\hskip 8mm}cccccccc}
\tablecaption{Two-sample KS Tests of \ac{cso} Redshifts as a Distinct Population}
\tablehead{Test&Complete&Sky& Flux Density&Frequency&KS&p-value & Significance\\
Number&Sample&Area&limit&GHz&statistic&&}
\startdata
1& PR+CJ1&$\delta > 35^\circ, |b|> 10^\circ$&0.7\,Jy &5\,GHz&0.34& $1.3 \times 10^{-1}$ &1.1$\sigma$ \\
2& PW&$\delta > 10^\circ, |b|> 10^\circ$&1.5\,Jy &2.7\,GHz&0.41& $3.1 \times 10^{-2}$ & 1.9$\sigma$\\
3&PWS&$10^\circ<\delta < 35^\circ, |b|> 10^\circ$&1.3\,Jy &5\,GHz&0.52& $1.2\times 10^{-1}$ & 1.3$\sigma$\\
4& PR+CJ1+PWS&-&-&-&-& $1.6 \times 10^{-2}$ &2.1$\sigma$ \\
\enddata
\tablecomments{Tests \#1 and \#3 are independent due to their the different sky areas. We can therefore, legitimately, multiply their p-values, which we do in Test \#4.    While the results of this redshift test do not rise to the $3\sigma$ level, and so cannot be considered to be compelling in and of themselves, they do strongly suggest a difference between most CSO 2s and the rest of the jetted-AGN population, and  are therefore supportive of the other tests we present.} 
\label{tab:ksredshift}
\end{deluxetable*}

We consider three populations of objects that are larger than CSO 2s and that might, therefore, be the populations that CSO 2s evolve into:
\vskip 6pt 
\noindent
(i) \ac{css} objects \citep{1982MNRAS.198..843P}, including the subclass of \acp{mso} which have sizes in the range 1\,kpc -- 20\,kpc \citep{1995AandA...302..317F} and R96. Note that MSOs have the same characteristics as CSO 2s apart from the size range.
\vskip 6pt 
\noindent
(ii) Fanaroff \& Riley Class~I jetted-\acp{agn} \citep{1974MNRAS.167P..31F}, which have sizes that range up to $\approx 1$ Mpc -- see Fig.  \ref{plt:histogramlargescale}(upper panel).
\vskip 6pt 
\noindent
(iii) Fanaroff \& Riley Class~II jetted-\acp{agn} \citep{1974MNRAS.167P..31F},  which also have sizes that range up to $\approx 1$ Mpc -- see Fig.  \ref{plt:histogramlargescale}(lower panel).

Note that \citet{1995AandA...302..317F} and R96 used an upper size limit of $15h^{-1}$\,kpc for \acp{mso}, where $H_o = 100 h \;
\mathrm{km\; s^{-1} \;\,Mpc^{-1}}$. For our adopted cosmology, this translates to 21\,kpc.  However, since the original choice of $15h^{-1}$\,kpc was chosen by \citet{1995AandA...302..317F} and R96 to be a convenient ``round number'',  we will follow that practice and use 20\,kpc  as the upper size limit of \acp{mso} in this study,   which also accords with the upper limit on MSO and CSS sizes adopted by OS21.

   The fact that the fraction of CSOs in complete samples is far too high for them  all to evolve into larger classes of radio sources was first discussed by R94, and has since been much studied (see OS21). Here we discuss the results for the hona fide CSO 2s in our complete samples. We see from  \cref{tab:csonumbers} that there are 19~CSO 2s and 43~\ac{css} objects in the combined PR+CJ1+PW complete sample. Note that MSOs are a subset of the CSS class. Thus the fraction of CSO 2s in the combined \ac{cso}+\ac{css} sample is $(30 \pm 8)\%$.   Assuming an upper limit on \ac{css} and \ac{mso} sizes of 20\,kpc,  on our hypothesis of constant speed of advance, the number of CSO 2s in complete samples of \ac{css} and \acp{mso} should be  $1/20=5\%$.  We therefore reject the hypothesis that a significant fraction  of CSO 2s  evolve into \ac{css}+\ac{mso} sources.

 The median size of the FR-I sources in the combined PR, CJ1 and PW samples shown in Fig.  \ref{plt:histogramlargescale} is 180 kpc.  This can be compared to the median size of the CSO 2s in these samples of 215 pc.  The ratio in sizes $\approx 837$, so that on the hypothesis of constant expansion speed we would expect there to be $\sim$15,907 FR Is, whereas there are 16 --- i.e., there are $\sim 990 \times$ fewer FR Is than expected.   Conversely, given the number of FR Is in these three complete samples, there are $\approx 990 \times$ more CSO 2s than expected.

 The median size of the FR-II sources in the combined PR, CJ1 and PW samples shown in Fig.  \ref{plt:histogramlargescale} is 305 kpc.  This can be compared to the median size of the CSO 2s in these samples of 215 pc.  The ratio in sizes $\approx 1420$, so that on the hypothesis of constant expansion speed we would expect there to be $\sim$27,000 FR IIs, whereas there are 77 --- i.e., there are $\sim 350 \times$ fewer FR IIs than expected. Conversely, given the number of FR IIs in these three complete samples, there are $\approx 350 \times$ more CSO 2s than expected. 

 We see therefore that the numbers of both FR~I and FR~II sources, relative to CSO 2s are far too small, by  factors of over 900 for the FR~Is and over 300 for the FR~IIs, for  CSO 2s to evolve into either FR~I or FR~II sources of comparable radio luminosity. At this flux density level the integrated number-flux density counts have a power-law slope of $-1.3$, so that the luminosity would have to drop by  factors of 200 and 90, respectively to accommodate this scenario for FR~I or FR~II objects.

 \citet{1991Natur.349..138R} have shown that there is a strong correlation between radio jet power and optical narrow line luminosity. Based on observations by \citet{1996ApJS..107..541L}, R96 showed that the narrow line luminosities of the CSO 2s J0111+3906, J0713+4349 and  J2355+4950 are about a factor 30 below that of typical FR-II galaxies, so that if CSO 2s are to evolve into FR~II galaxies, then their optical line luminosities must increase by about a factor 30 while their radio luminosities decrease by about a factor 35, which seems an unlikely scenario.
 It is interesting to note that R96 show that the jet power for J2355+4950, when corrected for the Hubble constant and different cosmologies, is $\sim 7 \times 10^{43}\; {\rm erg \, s^{-1}}$, and for J0111+3906, and J0713+4349 the similarly corrected jet powers $\sim 10^{45}\; {\rm erg \, s^{-1}}$, which may be compared to the range of jet powers in FR II sources of $\sim 10^{44}\; {\rm erg \, s^{-1}} - 10^{47}\; {\rm erg \, s^{-1}}$ (R96). Thus the jet powers of CSO 2s are similar to those of FR II objects, as is also the case regarding their luminosities. Given the agreement in narrow line luminosity between CSO 2s and FR~I galaxies, the possible evolutionary scenario from CSO 2s to FR~I galaxies may seem promising, but again, the numbers are off by over a factor 900.

We conclude on the basis of these fractions of CSO 2s in complete samples, that the vast majority ($\gtrsim 99\%$) of CSO 2s do not evolve into any of the above classes of larger jetted-\acp{agn}, and  therefore that they belong to a distinct class of jetted-AGN.

\begin{figure}[!t]
 \centering
 \includegraphics[width=0.6\columnwidth]{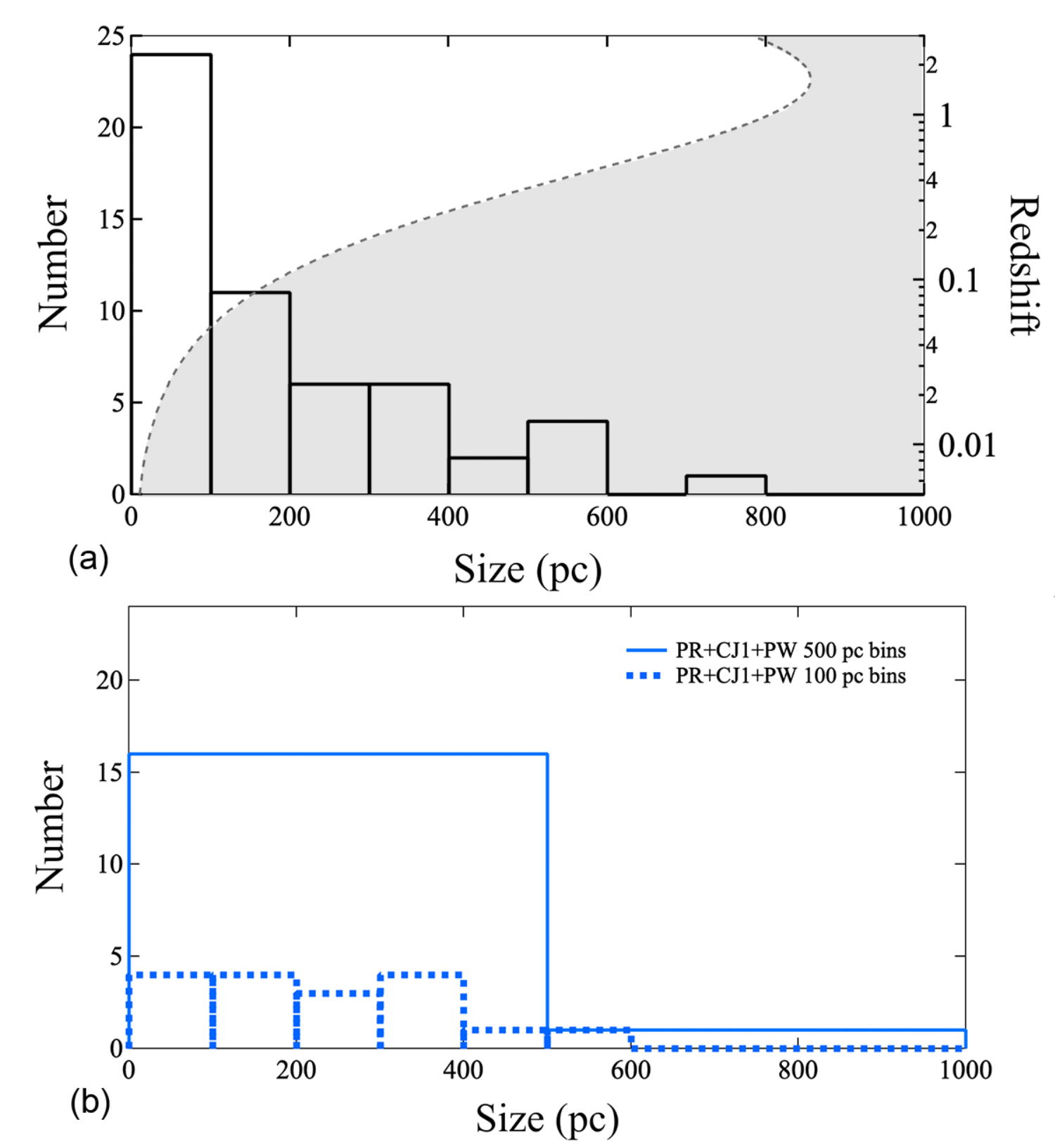}
 \caption{ The distributions in size of the bona fide CSOs over the whole \ac{cso} size range, from 0\,pc to 1\,kpc: (a) The heavy black boxes show the histogram of  the sample of 54~bona fide CSOs for which there are spectroscopic redshifts, with the numbers given on the left axis. The dashed curve, marking the border of the shaded region, shows the physical size corresponding to 100 milliarcseconds at the redshift indicated on the right-hand axis. For typical \ac{vlbi} observations at 5\,GHz and above, CSOs in the grayed region to the right of this curve would be hard to observe, so there is a strong selection effect that might account for the drop in numbers of bona fide CSOs with physical size. (b) The 17~bona fide CSO 2s with spectroscopic redshifts in the complete flux density-limited PR+CJ1+PW sample. Dotted curves show the data binned into 100 pc bins, while solid curves show the data binned into 500 pc bins.  The Kolmogorov-Smirnov and binomial tests both show that this distribution differs from a uniform distribution at the p-value $\sim1.7\times10^{-4}\;(3.6\sigma)$ level. While the observed uneven distribution could just be a result of small statistics, it would be foolish to ignore it, especially in light of the corroborating evidence from both the numbers and the redshift distributions. Nature often surprises us.}
 \label{plt:histograms}
\end{figure}

\section{The Redshift Distribution of CSO 2\lowercase{s} in Complete Samples}\label{sec:redshift}

An independent test of whether or not CSO 2s are drawn from the same population as the other jetted-\acp{agn} in our complete samples is provided through the redshift distribution. The redshifts are listed in Table \ref{tab:zandsize}.  The redshift distributions of the PR+CJ1 and PW complete samples, and their corresponding \ac{cso} distributions, are shown in \cref{plt:histogramredshift}.

We have carried out the \ac{ks} 2-sample test on the PR+CJ1 sample, the PW sample,  and the PWS sample, with the results given in the  four tests shown in \cref{tab:ksredshift}. The cumulative distributions corresponding to Tests \#1, \#2 and \#3, and their KS statistics, are shown in \cref{plt:KSredshift}~(a), (b) and~(c). In carrying out these tests we have removed the CSO 2s from the full samples.    The KS statistic is completely determined by the data, but the corresponding p-value depends on the assumptions made in integrating over the parent distribution \citep{1992nrfa.book.....P}. We verified that MATLAB and Numerical Recipes use the same formulae for determining the p-values.  For that reason we use the MATLAB p-values in deriving the significance levels in \cref{tab:ksredshift}.

\begin{figure*}[!t]
 \centering
 \includegraphics[width=1.0\linewidth]{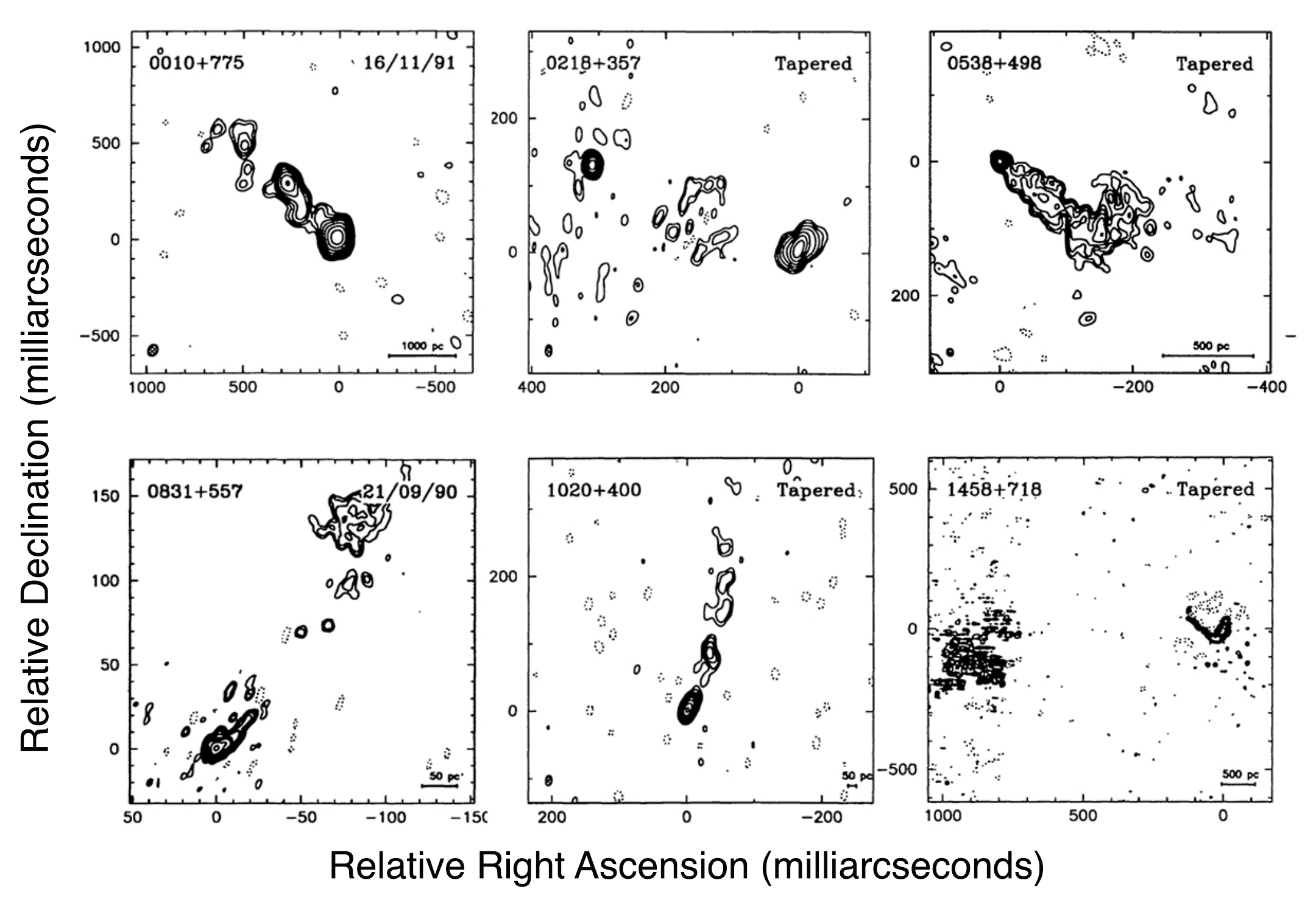}
 \caption{Demonstration that the complete samples studied in this paper are not restricted by the usual $\sim 100$~milliarcsecond field sizes typical of most \ac{vlbi} observations at 5\,GHz. Shown here are 1.7\,GHz \ac{vlbi} maps of six large angular scale compact AGN from the CJ1 complete sample survey \citep{1995ApJS...98....1P}, all of which have sizes $\gg 100$~milliarcseconds. Note that the structure of 1458+718 (J1459+7140, 3C~309.1) extends over 1\,arc second --- i.e., the map is ten times larger than the typical field of view of \ac{vlbi} maps at 5\,GHz or higher frequencies.}
 \label{plt:largeobjects}. 
\end{figure*}

The first two redshift distribution tests (\#s 1 and 2 in \cref{tab:ksredshift}) show that the probability of the hypothesis that the CSO 2s and non-CSO 2s are drawn from the same population is 0.13 for the PR+CJ1 sample;  and 0.03  for the  PW sample. 
If we look at the effectively complete PW subsample having S$_{\rm 5\,GHz} >1.3$\,Jy and at declination $\delta < 35^\circ$, which is independent of the PR+CJ1 sample in view of the mutually exclusive declination limits, we see that the  probability is ~0.12. Since these are independent samples, we may legitimately multiply the p-values of Tests~\#1 and~\#3, which yields a probability of $1.6\times 10^{-2}$, which is significant at the $2.1\sigma$ level. While not at the $3\sigma$ level, these statistics nevertheless provide some independent evidence that CSO 2s are drawn from a different population compared to that of the other jetted-\acp{agn} in these complete samples. 

This result, which is seen clearly in the redshift distributions shown in \cref{plt:histogramredshift}, is interesting.  If correct, it suggests that CSO 2s only started forming in significant numbers towards the end of the epoch of maximum galaxy and star formation: The lookback time to the peak in the cosmic star formation rate is $\sim 8$ billion years \citep{2020ARA&A..58..661F}, which is close to the lookback time to  redshift $z \approx 0.9$, when CSO 2s started to appear in significant numbers, as can be seen in \cref{plt:histogramredshift}(a) and (b). The peak \ac{sfr} occurs from $z \approx 1$ to $z \approx 2$, with the peak \ac{smbh} formation rate \citep{2020ARA&A..58..157T} peaking slightly after the peak \ac{sfr}.

Thus, a possible explanation of the origin of CSO 2s is that quiescent \acp{smbh} form CSO 2s  by single star capture, and so become significant around $z\sim 1$, when the numbers of both stars and \acp{smbh} in the universe reach a maximum.  We give a detailed discussion of this hypothesis in Paper~3.

However  results that are significant only at the $\sim  2 \sigma$ level often disappear with the advent of more data, and  this particular  apparent difference between CSO 2s and other jetted-\ac{agn} may disappear as more bona fide CSO 2s are accrued through new and deeper complete samples.  An important point that should be mentioned here is that there are 14 bona fide CSOs in the  VLBA Imaging and Polarization Survey (VIPS) of flat spectrum ($\alpha \geq -0.5)$)  sources \citep{2007ApJ...658..203H} that are not in the complete PR+CJ1+PW samples, and of these only one has redshift greater than 1. Thus, extending the luminosity function almost an order of magnitude deeper appears not to change our findings in this section. This gives us  confidence that this preliminary $2.1 \sigma$ result will be greatly strengthened, when we are able to add the steep spectrum counterparts to the VIPS survey to make this a complete sample, as we are now engaged in doing.

As a distinct population, and recalling that these are all ``short-lived'' but not all ``young'' sources, it will be of great interest to investigate whether \ac{cso}~2s show the same strong cosmological evolution as do  both high-luminosity  extended steep spectrum sources and compact flat spectrum sources \citep{1981MNRAS.196..611P}, but this is beyond the scope of the present paper.

\section{The Size Distribution of CSO\lowercase{s}}\label{sec:size}

Our third independent test of the hypothesis that CSO 2s form a distinct class of jetted-\acp{agn} is based on the size distribution of CSO 2s. This test is more complex and more subject to selection effects than the tests of the previous two sections.  Selection effects are particularly strong when it comes to consideration of the observed distribution of \ac{cso} sizes, so we discuss first the effectiveness of our approach in dealing with these selection effects, in order to give the reader some confidence in the statistical robustness of our results.

\subsection{The Efficacy of Complete Samples in Dealing with the CSO Size Distribution Selection Effects}\label{sec:efficacy}

The distribution of the physical sizes of the 54~bona fide CSO 2s, out of our sample of~79, for which we have spectroscopic redshifts is shown in \cref{plt:histograms}~(a). It shows a strong cutoff well below 1\,kpc. However, one has to bear in mind that this sample of 54~bona fide CSO 2s is a heterogeneous sample gleaned from the literature, and is subject to selection effects. We therefore have to consider carefully whether these selection effects can be eliminated in complete sub-samples of our 54~bona fide CSO 2s.

\subsubsection{The \texorpdfstring{$\sim 100$}{100} Milliarcsecond Selection Effect}\label{sec:typical}

The first selection effect we consider comes about because the largest angular size that is
measured in most cm-wavelength \ac{vlbi} maps $\sim 100$~milliarcseconds. In \cref{plt:histograms}~(a) we show the upper size cutoff this would impose as a function of redshift. Only CSOs to the left of this curve have angular sizes less than 100~milliarcseconds at the corresponding redshift. Clearly this could well impose a strong selection effect on the observed size distribution of CSOs.

On the face of it, it might appear that this selection effect alone is so strong that the true size distribution of CSOs is impossible to determine from these data. Fortunately this is not the case because one can observe complete samples in which one knows the sizes of all the objects in the sample, and if some objects are too large for \ac{vlbi} mapping at cm wavelengths they can be observed at longer wavelengths, where the $\sim 100\,$mas limit does not apply. We have availed ourselves fully of this strategy: In addition to the observations of compact objects in the PR+CJ1 complete samples at 5\,GHz \citep{1988ApJ...328..114P,1995ApJS...99..297X},   all of these objects were observed at 1.66\,GHz \citep{1995ApJS...98....1P,1995ApJS...99..297X}.
In \cref{plt:largeobjects} we show examples of six AGN from  the CJ1 complete sample with sizes far exceeding the 100\,mas angular size limitation of regular \ac{vlbi} at 5\,GHz and above. As can be seen here, even objects as large as 1 arcsecond were mapped in this survey. This is one of two reasons we can be confident that we have not missed any large CSOs in these complete samples. The other reason is that the large-, by which we mean ($\gtrsim$ 1 arcsec), -and small-scale radio structures of {\it all} of these objects are known. In  \cref{tab:samples} in the Appendix we list references which present the relevant maps of all 281 objects in the PR+CJ1+PW samples.

\subsubsection{Spectral Shape Selection Effect}\label{sec:spectra}

Many CSOs are peaked spectrum (PS) sources\footnote{In this paper we follow the lead of  \citet{2021AandARv..29....3O} in their comprehensive review of peaked spectrum sources, and refer to GPS and MHz peaked spectrum sources as Peaked Spectrum (PS) sources}. Thus in a sample of CSOs selected at a single frequency, we will clearly include all of the sources that peak at that frequency down to the flux density limit.  However, for sources that peak at frequencies significantly higher or lower than the selection frequency, the sample will exclude an increasing number of the CSOs as the separation between the peak frequency and the selection frequency increases.   In this study, we therefore consider not only the PR+CJ1 and PW samples, selected at 5 GHz and 2.7 GHz, but we also consider the GLEAM sample, observed using 20 simultaneous flux density measurements spanning frequencies between 72 MHz and 231 MHz,  the 3CRR sample selected at 178 MHz, and the Jodrell Bank 966 MHz sample.

\begin{figure}[!t]
 \centering
 \includegraphics[width=1.0\linewidth]{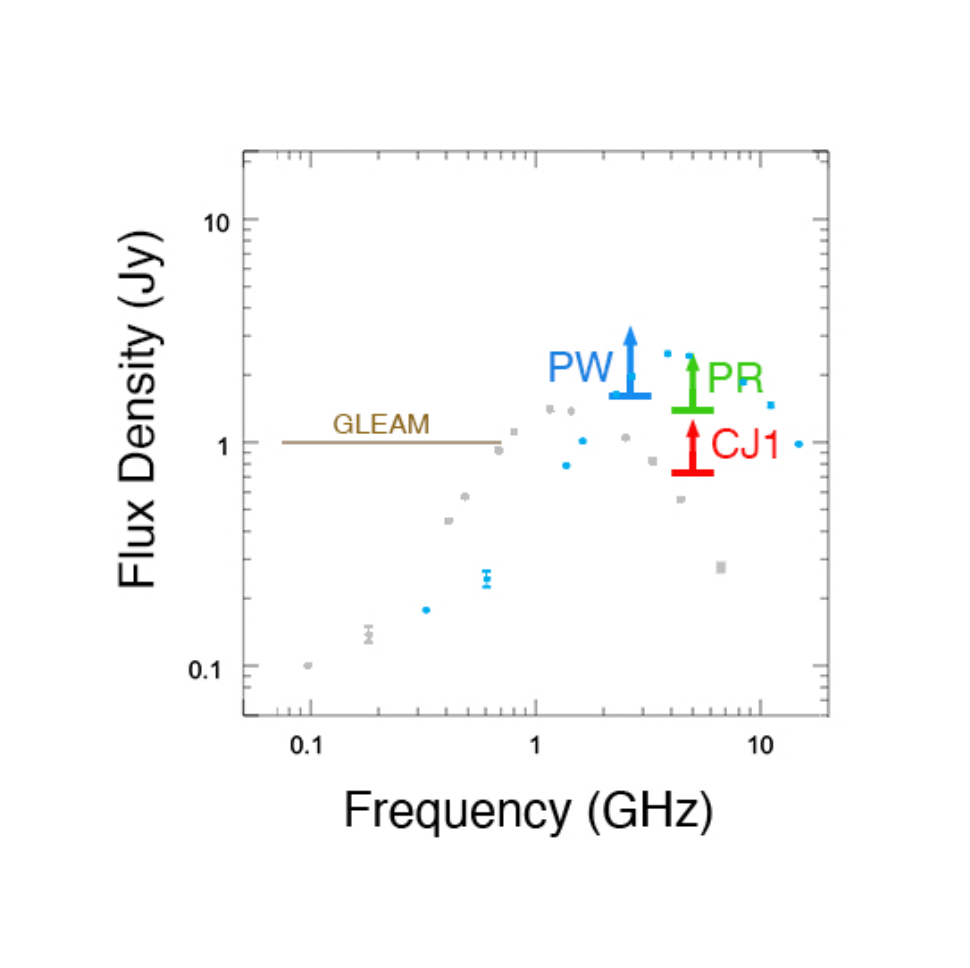}
 \caption{CSO 2s that might be missed in the PR, CJ1, and PW complete samples due to spectral effects: The blue, green and red arrows indicate the selection frequency and limiting flux densities of the PW, PR, and CJ1 samples, respectively. The horizontal brown line indicates the flux density limit of the GLEAM sample. The blue and gray points show the observed spectrum of OQ~208 \citep{1997A&A...318..376S}, and a shifted spectrum of  OQ~208, respectively (see text).}
 \label{plt:OQ208}
\end{figure}

The situation is illustrated in \cref{plt:OQ208}. The blue points show the observed radio spectrum of OQ~208 (J1407+2827) \citep{1997A&A...318..376S}, which has one of the narrowest, most sharply peaked, spectra amongst the bona fide CSO 2s. The gray points illustrate an object with the same spectral shape as OQ~208, but with the maximum shifted from 5\,GHz down to 1\,GHz, and the peak  flux density shifted down to 1.3\,Jy. This is the point where the object would drop below the GLEAM 1\,Jy limit \citep{2017ApJ...836..174C}, and the CJ1 700\,mJy limit. Because of the drop-off in flux density, relative to the peak,  at both higher and lower frequencies,  such an object would not be included in  the PW, PR, 
 CJ1 or GLEAM samples. Objects of this type with peak flux densities greater than 1.3\,Jy would, however, be included in the GLEAM and CJ1 samples, whose limiting flux densities are indicated by the horizontal brown line in \cref{plt:OQ208}, and the red horizontal bar, respectively.
  We  therefore consider next what is known about the population of objects that peak at frequencies $\lesssim 1$ GHz.

\begin{figure}[!t]
 \centering
 \includegraphics[width=1.0\linewidth]{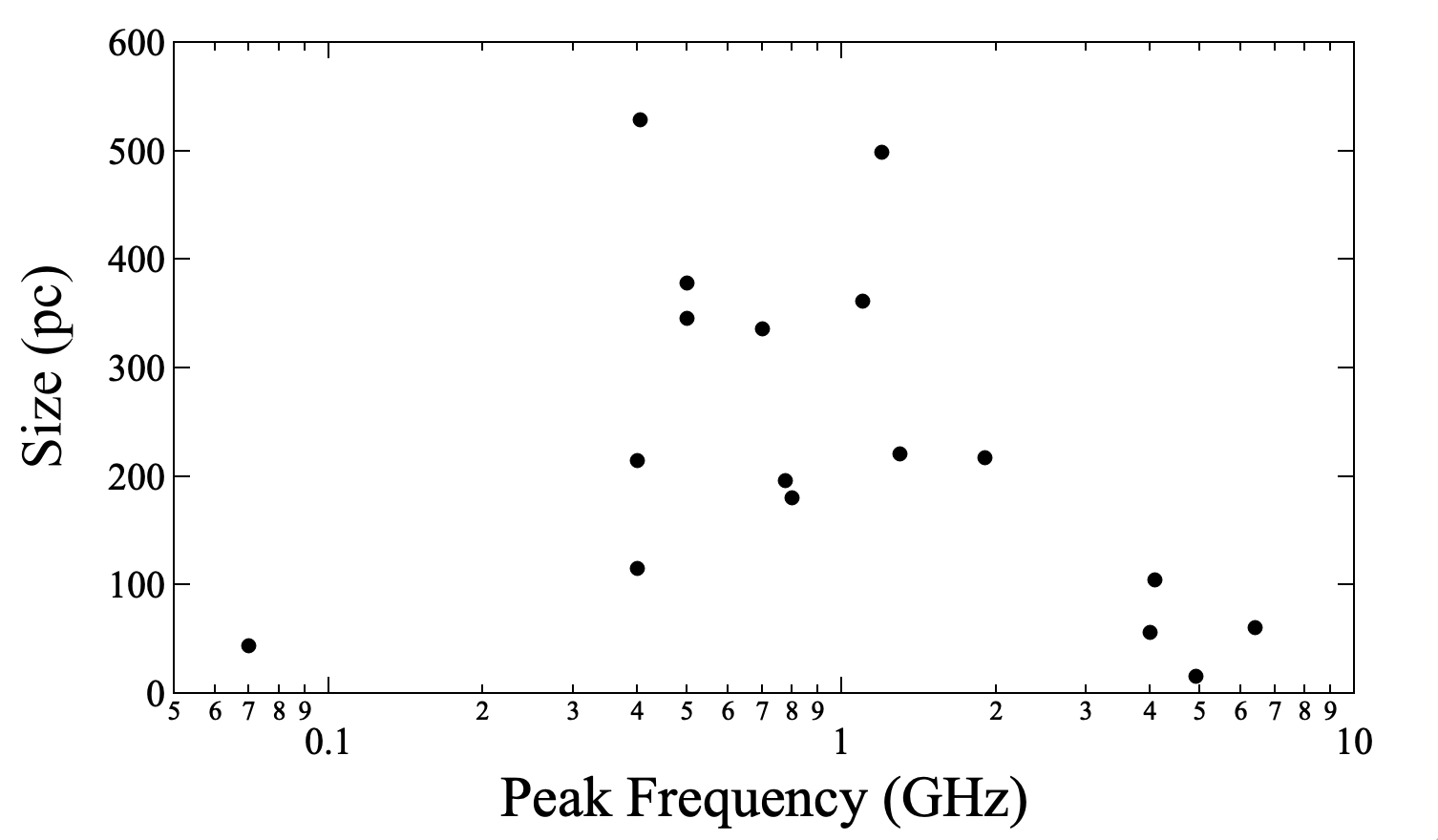}
 \caption{The relationship between size and peak frequency for the CSO 2s in the PR+CJ1+PW complete sample.}
 \label{plt:siszevspk}
\end{figure}

The GLEAM survey  covers 72 MHz -- 700 MHz, and is concentrated in the southern hemisphere, and so overlaps only part of the sky area covered by the PR+CJ1+PW sample, but it complements the higher frequency \ac{vlbi} surveys that have studied complete samples because of its lower frequency.   505 of the 11,400 sources in the complete 1\,Jy GLEAM sample (($4.4 \pm 0.2)$ \%) are peaked-spectrum objects \citep{2017ApJ...836..174C}, with peak flux densities above 1\,Jy in the 72\,MHz -- 700\,MHz range,  and so must also be compact \citep{2021AN....342.1185R}.    Similarly, in a  LOFAR study of  northern radio sources at 150 MHz, \citet{2022A&A...668A.186S} found  373 PS sources and concluded that  $\sim2.5\%$ of  sources in  complete samples around 150 MHz are PS sources. We note the similarity  in the fractions of PS sources identified in the  LOFAR (150 MHz), GLEAM (72 MHz - 700 MHz), and the fraction of PW+CJ1+PW (2.7 GHz and 5 GHz) CSO 2s samples, which are $\sim 2.5\%$, $\sim 4.4\%$, and $\sim 6.8\%$, respectively. It is possible that a significant fraction of the PS population in both the LOFAR and GLEAM surveys are CSO 2s, and thus that   there is a significant population of CSO 2s extending to below 100 MHz. In this case we could well be missing CSO 2s that could fall into in the 500 pc -- 1 kpc size range. In Fig. \ref{plt:siszevspk} we show the sizes of the PR+CJ1+PW CSO 2s plotted against peak frequency.  It is interesting to note that almost half of these objects have peak frequencies in the range covered by the GLEAM survey, even though they were selected at 2.7 GH and/or 5 GHZ.  It is also interesting to note that, apart from the small fraction ($< 25\%$) that has peak frequencies above 3 GHz, there is no clear dependence of size on frequency.

\subsection{A Spectral Shape Lacuna}\label{sec:3CRPW}

As we have seen in previous sections, we are only considering the 17 bona fide CSO 2s in the PR+CJ2+PW complete sample with known spectroscopic redshifts, and there are only two bona fide CSOs in these complete samples  without a spectroscopic redshift. The PR+CJ2+PW complete sample, excluding M82, consists of the 281  sources listed in Table \ref{tab:samples}, including M82.  In \ref{sec:spectra} we discussed a spectral shape selection effect that can be affecting this sample. 
The GLEAM survey detected 11,400 sources with flux densities greater than 1 Jy between 70 MHz and 700 MHz.  Of these 505 are PS sources.   In order to double the numbers of CSO 2s, and hence potentially to have a strong effect on any statistical tests of the size distribution of CSO 2s, there would need to be $\approx$17 bona fide CSO 2s in the GLEAM sample.  Thus only a small fraction $\sim 3 \%$ (17/505) of the PS sources in the GLEAM would need to be CSO 2s in order potentially to have a significant impact on the statistics.  So this is a lacuna that has to be addressed in any size tests.

In the next three subsections we advance two independent arguments to address this lacuna and we suggest a test that could fill the lacuna, but which requires more observations and is  therefore beyond the scope of the present paper. 

\subsubsection{The Range of Peak Frequencies in Our Sample}\label{sec:rangepeak}
  
In Paper~1, Fig. 6, we have plotted the range of peak frequencies observed, and it can be seen  that the peak frequencies range from below 80 MHz to  $\sim 10$ GHz.  The same is true of the objects in our combined PR+CJ1+PW sample -  the lowest peak is at 70 MHz and the highest peak is at  $6.4$ GHz, and the distribution of the peaks is roughly uniform between these extremes.  

 Thus the selection procedure of the complete PR+CJ1+PW sample and our bona fide CSO identification method do not appear to have created a bias against CSO 2s peaking anywhere within this range. However, while the (rarer) flat-spectrum CSO 2s will not suffer from the spectral selection biases described earlier, some peaked-spectrum CSO 2s could be excluded from the sample for certain redshift ranges.  This could, therefore, influence the size distribution of the observed CSO 2s in the PR+CJ1+PW samples, particularly if CSO intrinsic size is related to peak frequency and/or luminosity.

 In the next two subsections we give an argument that shows that spectral shape selection effects are unlikely to have biased the size distribution of the CSO 2s in the PR+CJ1+PW sample.

\subsubsection{The 3CRR and PW CSS Double Sample}\label{sec:lowfreq}

In addition to our complete samples of CSO 2s, described in  the previous sections, there is one other relevant sample of CSO 2s and MSOs that has been studied extensively by the Bologna Group (BG), the key results of which are given in a series of papers \citep{1985AandA...143..292F,1990AandA...231..333F,1991MNRAS.250..225S,1995AandA...302..317F,1995AA...295...27D,2013MNRAS.433..147D,2021MNRAS.504.2312D}. The BG identified 32 double-lobed CSS objects, given in Table 1 of \citet{1995AandA...302..317F}, in their sample drawn from the 3CRR \citep{1983MNRAS.204..151L} and the PW samples. They subsequently added one double-lobed source (1819+396 = 4C $+$39.56) that they had previously missed \citep{2021MNRAS.504.2312D}, bringing the total of double-lobed CSS sources in the BG sample to 33. Given that these are CSS objects, they excluded flat spectrum objects with $\alpha>-0.5$.

 This spectral filter  against flat spectrum sources ($\alpha > -0.5$), as applied to the 3CRR sample, which has a limiting flux density of 10 Jy at 178 MHz, excludes sources
 brighter than 2.57 Jy at 2.7 GHz.  Since these are greater than the flux density limit of the PW complete sample  (1.5 Jy at 2.7 GHz), any such  object can be included in the BG study by adding the flat spectrum PW CSO 2s to the steep spectrum CSO 2s detected by the BG, in order to get the total of both flat and steep spectrum compact doubles, including CSO 2s, in the 3CRR and PW samples (note that there are no flat spectrum doubles of size greater than 1 kpc in PW).   There are four BG CSOs in our sample of 79 bona fide CSOs listed in Paper 1. All four of these BG CSOs are already in our complete sample of CSOs  in the PR+CJ1+PW complete samples.

 \begin{figure*}[!t]
    \centering
    \includegraphics[width=1.0\linewidth]{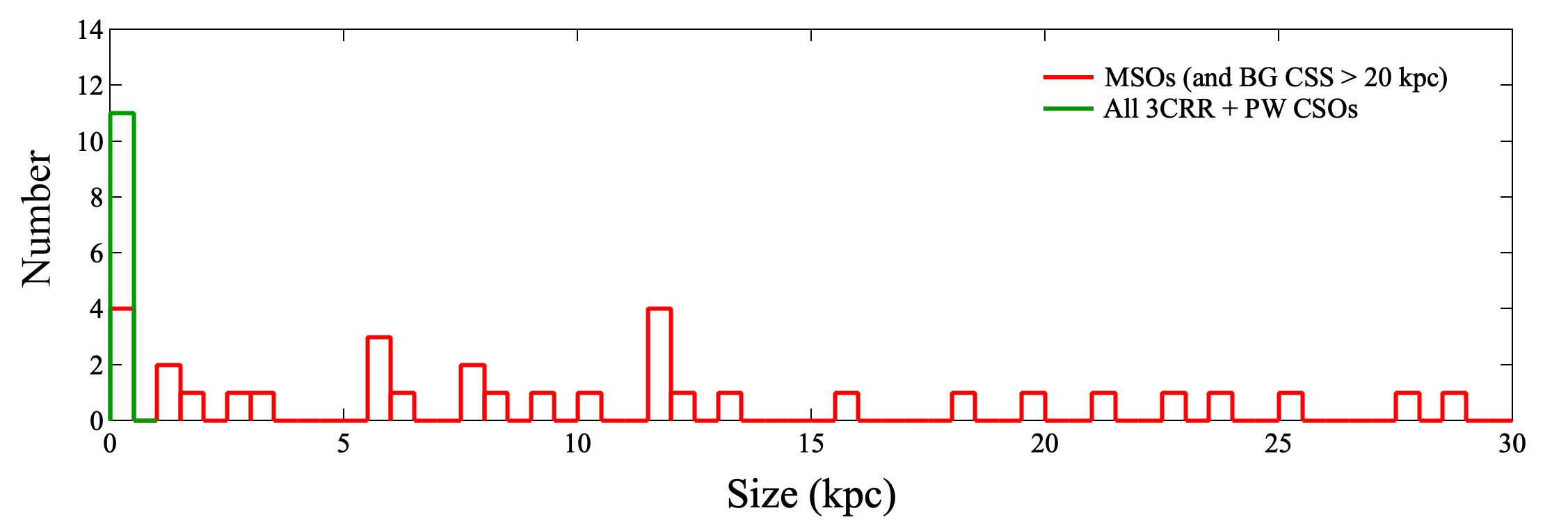}
    \caption{ The distributions of the bona fide CSO 2s and MSOs as a function of projected physical size in the BG sample. The red distribution shows the BG CSO 2s+MSOs+CSS$>$20 kpc objects.  The green distribution shows the BG+PW CSO 2s. }
    \label{plt:bghistograms}
\end{figure*}

 We therefore compensate for the spectral index limit in BG sample by adding the flat spectrum PW bona fide CSO 2s that were excluded from the BG sample by the spectral index cutoff at $\alpha =-0.5$ to the BG sample of CSO 2s, thereby  making this into a complete sample of  3CRR+PW compact double sources, and bringing the total including the PW flat spectrum CSO 2s to 40.   In order to apply the same largest angular size filter as that used in Paper~1, we have re-measured the largest angular sizes of the 33 BG CSS double sources at the lowest frequency at which high-quality maps are available in the BG group's publications listed above. The results are shown in Fig. \ref{plt:bghistograms}, together with the 7 flat spectrum PW bona fide CSO 2s that we have added.

We find that 27 of the 33 objects in the BG sample fit the CSO$+$MSO criteria, with four of the objects being bona fide CSO 2s and the remaining 23 objects being MSOs. The 6 remaining objects  all have largest  projected physical sizes greater than 20 kpc, based on our measured largest angular sizes.  When we add the flat spectrum objects from the PW sample, the number of CSO 2s increases from 4 to 11, as shown in Fig. \ref{plt:bghistograms}. 

 The four CSO 2s in the BG CSS sample all have sizes between 300 pc and 500 pc.  These PS CSO 2s have spectral turnovers that are almost certainly due to synchrotron self absorption as is shown in the paper by \citet{1977MNRAS.180..539S}, who showed that
 the equipartition angular size
$\psi_{_{\rm eq}} \propto  S^{8 \over 17}  \nu^{-{{35+ 2\alpha} \over 34}}   (1+z)^{{15-2\alpha} \over 34}$.  Thus fainter objects that show spectral peaks at higher frequencies will be smaller than the four CSS CSO objects in the BG sample. It therefore is unlikely that there will be a significant number of CSS CSO 2s with sizes in the 500 pc to 1 kpc range.  We return to this point in \S \ref{sec:compsizes}.

\begin{figure*}[!t]
 \centering
 \includegraphics[width=1.0\linewidth]{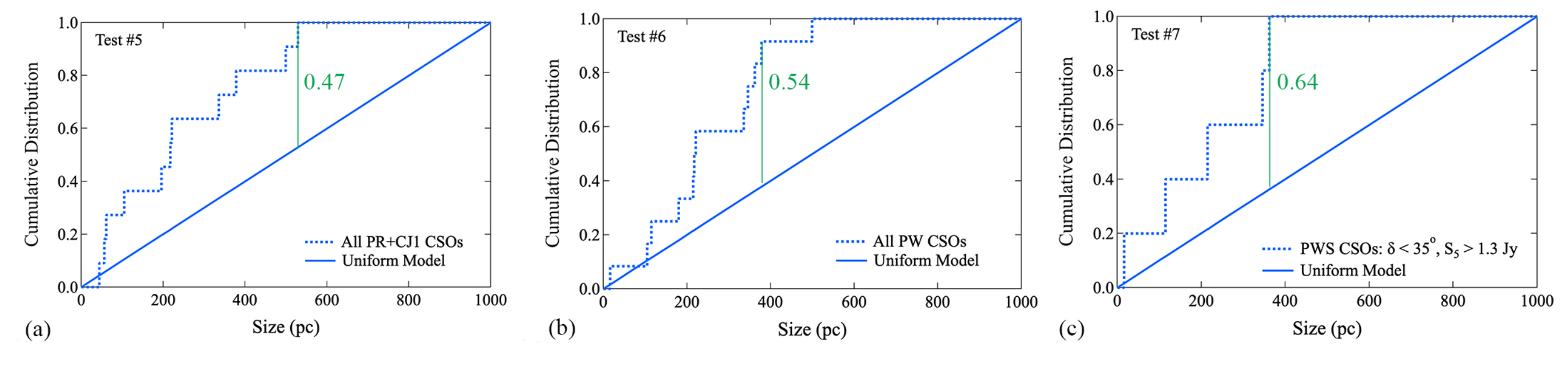}
 \caption{KS Tests on the  size distributions of the bona fide CSO 2s in the PR+CJ1 and PW samples. (a), (b) and (c): comparison of the CSO 2 cumulative size distributions {\it vs.}  the uniform model for the  PR+CJ1, PW, and PWS samples, respectively. The green bars indicate the maximum differences in the cumulative distributions, corresponding to the values of the KS statistic given by the numbers in green. The corresponding p-values are listed in \cref{tab:kssize}. }
 \label{plt:KSsize}. 
\end{figure*}

\begin{deluxetable*}{c@{\hskip 8mm}cccccccc}
\tablecaption{One-sample KS Tests Against a Uniform Distribution  of CSO 2 Sizes\label{tab:kssize}}
\tablehead{Test&Complete&Sky& Flux Density&Frequency&KS&p-value & Significance\\
Number&Sample&Area&limit&GHz&statistic&&}
\startdata
5& PR+CJ1&$\delta > 35^\circ, |b|> 10^\circ$&0.7\,Jy &5\,GHz&0.47& $9.3 \times 10^{-3}$ &2.4$\sigma$\\
6&PW&$\delta > 10^\circ, |b|> 10^\circ$&1.5\,Jy&2.7\,GHz&0.54& $8.7 \times 10^{-4}$ & 3.1$\sigma$\\
7& PWS&$10^\circ<\delta < 35^\circ, |b|> 10^\circ$&1.3\,Jy&5\,GHz &0.64& $1.7 \times 10^{-2}$ & 2.1$\sigma$\\
8& PR+CJ1+PWS&-&- &-&-& $1.6 \times 10^{-4}$ & 3.6$\sigma$\\
\enddata
\tablecomments{Tests of the observed CSO 2 size distribution compared to a uniform size distribution. The PWS sample is the effectively complete subsample of the PW sample having $10^\circ<\delta < 35^\circ, |b|> 10^\circ$ and  S$_{\rm 5\,GHz} \ge 1.3 $\,Jy (see text).   }
\end{deluxetable*}

\begin{deluxetable*}{c@{\hskip 8mm}cccccccc}
\tablecaption{Binomial Tests of Significance Levels of \ac{cso} Size Distributions in Complete Samples}
\tablehead{Test&Complete&Sky& $\alpha$& N&N&p-value & Significance\\
Number&Sample&Area&limit&$0-500$\,pc&500\,pc-1\,kpc&&}
\startdata
9& PR+CJ1&$\delta > 35^\circ, |b|> 10^\circ$&- & 11& 1& $5.4 \times 10^{-3}$ & 2.6$\sigma$\\
10& PWS&$\delta < 35^\circ, |b|> 10^\circ$&-&6&0& $3.1 \times 10^{-2}$ & 1.9$\sigma$\\
11& PR+CJ1+PWS&-&- & -& -& $1.7 \times 10^{-4}$ & 3.6$\sigma$\\
\enddata
\tablecomments{ The PR+CJ1 and PWS ($\delta<35^\circ, \; {\rm S_5>1.3 \,Jy}$) samples are independent, so we have multiplied their p-values in Test 11 (see text). }
\label{tab:histograms}
\end{deluxetable*}

\subsubsection{The Jodrell Bank 966 MHz Sample}\label{sec:jodrell}

Since the 3CRR sample is complete down to 10 Jy at 178 MHz \citep{1983MNRAS.204..151L},  for comparison with the other samples in this study it would be helpful to have a low frequency sample complete down to $\sim 1$ Jy. Fortunately, such a sample exists for which the radio structures of over 98\% of the objects are known.

Referring back to Fig. \ref{plt:OQ208} and the objects in the lacuna illustrated by the gray spectrum. We can define a complete sample drawn from the Jodrell Bank 966 MHz survey \citep{1977MmRAS..84....1C}, which produced  a radio catalogue and measured arcsec-level positions for the majority of its sources.  We have selected a  sub-sample, consisting of 169 of the strongest sources (S$_{0.966} > 1.5$ Jy)  from \citet{1977MmRAS..84....1C}.   This sub-sample is unbiased, and while the full survey is  not strictly complete due to confusion issues, these apply only at  low flux density levels, and thus the sub-sample  that we have selected is not affected by confusion. We will refer to this unbiased sub-sample of 169 objects as the ``JBS'' sample.

We classified 74 of the JBS sample in the filtering process we carried out in selecting our bona fide CSO 2s described in Paper~1.  We identified six of them as bona fide CSO 2s.

We have extracted VLASS cutout images of all 169 JBS objects using the CIRADA cutout server\footnote{http://cutouts.cirada.ca}, and we found only 17 of them to be unresolved, with largest angular size $<3$ arc sec, and hence possible CSO 2s.  Of these 17 compact objects, two are MOJAVE ``core-jet'' objects, and  one is a 2 arc second double. Thus there are 14 possible CSO 2s in the 966 MHz JBS sample in addition  to the 6 bona fide CSO 2s we have already identified.  We are engaged in obtaining VLBI observations of these 14 objects.

\subsection{Statistical Analysis of CSO Sizes in the Complete Samples}\label{sec:compsizes}
  As we have seen, all of the bona fide CSOs in the complete PR, CJ1 and PW samples are \ac{cso}~2s --- i.e., as discussed in Paper~1, they are edge-brightened, high-luminosity objects. This is a selection effect resulting from the flux density limits in the complete samples.
 The size distribution of the  CSO 2s in the  PR+CJ1+PW  complete sample is shown in \cref{plt:histograms}(b), binned into 100 pc and 500 pc intervals.

Using the CSO 2 size distributions in the PR+CJ1 and PW complete samples, we have carried out two sets of statistical tests of the hypothesis that CSO 2s are uniformly distributed in size between 0\,pc and 1000\,pc, as would be expected on the hypothesis that the speed of advance is constant: (i) a set of \ac{ks} 1-sample tests, which yield the cumulative distributions shown in \cref{plt:KSsize}~(a), (b) and~(c) and the p-values shown in Tests \#5 - \#8 in \cref{tab:kssize}; and (ii) binomial tests in which we divided the CSO 2s into two size bins, from 0\,pc to 500\,pc, and from 500\,pc to 1000\,pc, which yield the results shown in \cref{tab:histograms}.

\begin{figure*}[!t]
 \centering
 \includegraphics[width=1.0\linewidth]{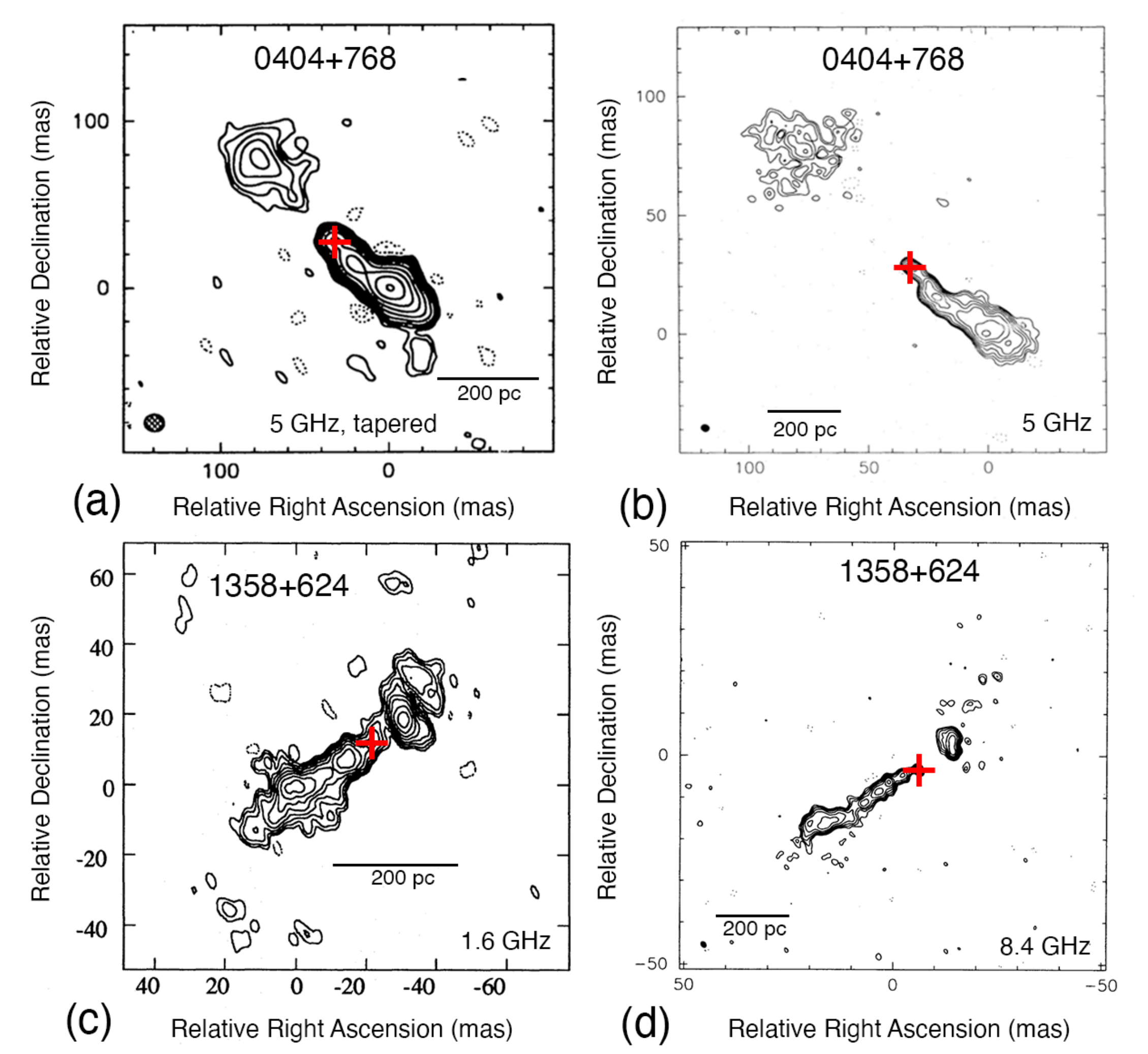}
 \caption{Example of an \ac{mso} and a \ac{cso}, with similar morphologies.  The MSO:(a) from \citet{1995ApJS...99..297X}, and (b) from \citet{1996ApJ...463...95T}:  0404+768 (J0410+7656), which these authors had classified as a \ac{cso}. The CSO:  from \citet{1995AA...295...27D}, and (d) from \citet{1996ApJ...463...95T}:  1358+624 (J1400+6210). Some CSO 2s must evolve into \acp{mso}. and 1358+624 may be just such a case. The red cross marks the location of the core in each map.}
 \label{plt:csomso}. 
\end{figure*}

We consider first the \ac{ks} tests shown in \cref{tab:kssize}.  We see there that the uniform hypothesis is rejected by the PR+CJ1 \ac{cso} sample at the $9.3 \times 10^{-3}$ probability level, and by the PW \ac{cso} sample at the $8.7 \times 10^{-4}$ probability level. The independent, effectively complete, PW \ac{cso} sample below declination $35^\circ$ (Test \#7) rejects the uniform hypothesis at the $1.7 \times 10^{-2}$ level.

 Tests \#5 and \#7 are independent -- note the different sky areas -- and at the same observing frequency. We can therefore, legitimately, multiply their p-values. This gives Test \#8, which rejects the uniform size distribution hypothesis at the p-value $1.6 \times 10^{-4}$, or $3.6 \sigma$ significance level.

The results of the binomial tests, shown in \cref{tab:histograms}, show that by similarly combining the PR+CJ1 \ac{cso} sample (Test \#9)  with the independent PWS sample (Test \#10), as shown in Test \#11, the uniform hypothesis is rejected at the $1.7\times 10^{-4}$ ($3.6 \sigma$) level.

In our view, these tests on complete samples constitute compelling evidence that the size distribution of CSO 2s cuts off sharply at $\approx 500$ pc, which is significantly below the 1\,kpc size limit imposed by the defining criteria of CSOs.   As described in \S \ref{sec:jodrell}, the existence of this sharp cutoff can be tested, for example, with MERLIN and VLBI observations of the 14 potential CSO 2s in the complete JBS sample.

Clearly {\it some} CSO 2s must grow to larger sizes in order to produce \acp{mso}. An example of a \ac{cso} and \ac{mso} with remarkably similar morphologies is shown in \cref{plt:csomso}, which illustrates this point, especially since 0404+768 (J0410+7656) was originally classified as a \ac{cso}, but fails the  size cutoff (by $\sim$20\%). It is clear that  the majority of CSO 2s do not grow much above 500\,pc in projected size.

\section{Results and Tests from This Paper}
\label{sec:major}
The results and tests that have come out of this study are listed below.

\subsection{Results}
\label{sec:results}

The number fraction, redshift, and size statistics discussed in the previous three sections and presented in \cref{tab:ksredshift,tab:kssize,tab:histograms} demonstrate that (i) most CSO 2s are drawn from a population of jetted-\acp{agn} that is distinct from other jetted-\acp{agn}, and (ii) the size distribution of CSO 2s cuts off sharply at $\approx 500$ pc, a finding that is verifiable (see \S \ref{sec:jodrell}).

These are  significant findings. They show that there has to be something fundamentally different between CSO 2s and the larger sources. While both must be driven by the same type of central engine, since both are producing high-luminosity relativistic jets, there must be some fundamental difference between them to produce two such different outcomes -- one with a size cutoff around 500 pc and the other with a size cutoff $\sim 200 \times$ larger.  

One might think, for example, that the cutoff could be explained simply by random episodic fuelling. But how would random episodic fuelling produce a sharp cutoff? Random episodic fuelling would produce a uniform distribution. There has to be another explanation for the cutoff, such as, for example, an upper limit on the size of the fuel packages,  a change in the jet environment that leads the jets to fade  beyond a certain distance from the central engine, or a mechanism associated with the accretion disk that limits the energy of the jet.

\subsection{Tests}
\label{sec:tests}

The major tests resulting from this paper are the following:
\vskip 6pt
\noindent
1. We will follow-up the JBS sample with MERLIN and  VLBA observations in order to test the sharp cutoff in size we have seen in the PR+CJ1+PW complete sample.
\vskip 6pt
\noindent
2. We have undertaken a program to increase the number of CSO 2s in complete samples by a factor of at least 3, to $\sim 50$ by carrying out a VLBI survey of $\sim 332$ steep spectrum sources that complement the VIPS flat 
 spectrum survey, thus converting VIPS into a complete  sample.   This will further test the cutoff in CSO 2 size distribution.
\vskip 6pt
\noindent
3. Structural studies of the PS objects identified in the LOFAR and GLEAM surveys would be well worth doing.

\section{Discussion}\label{sec:discussion}

Although the selection effects inherent in our literature search for CSOs are significant, we have shown that, with careful use of complete samples, and using the 17~CSO 2s in the complete PR, CJ1, and PW samples for which we have spectroscopic redshifts, it is possible to carry out a series of rigorous statistical tests that provide what we regard as compelling evidence that   the vast majority of    CSO 2s constitute a population of jetted-\acp{agn} that is distinct from, and therefore requires a separate origin to, the larger classes of jetted-\acp{agn}, such as \ac{css} sources,  \acp{mso}, and FR~I and FR~II objects.  

The physical size cutoff is clearly telling us something important about this class of jetted-\acp{agn}. The scenario that produces   almost all  CSO 2s must be different in some important way from that which produces the larger symmetric radio sources. We return to this discussion of the origins of CSO 2s in Paper~3.  

It should be clear, therefore, that, (i) because the observed emission regions in these objects are not significantly relativistically boosted towards the observer, thereby making it possible to determine their detailed physical properties, and (ii)  most CSO 2s  belong to a distinct class of jetted-\acp{agn},   these  CSO 2s provide a unique  time domain plus structural window on the central engines of jetted-\acp{agn} and the supermassive black holes that drive them.


\begin{acknowledgments}

We thank John Peacock for useful discussions.  We thank the reviewer of this paper for many helpful suggestions that have clarified several important aspects of this work. We are grateful for the use of the CATS database of \citet{2005BSAO...58..118V}, of the Special Astrophysical Observatory. 
This research has made use of NASA’s Astrophysics Data System Bibliographic Services.
This research has made use of the NASA/IPAC Extragalactic Database (NED) which is operated by the Jet Propulsion Laboratory, California Institute of Technology, under contract with the National Aeronautics and Space Administration.
This research has made use of data from the OVRO 40-m monitoring program (Richards, J. L. et al. 2011, ApJS, 194, 29), supported by private funding from the California Insitute of Technology and the Max Planck Institute for Radio Astronomy, and by NASA grants NNX08AW31G, NNX11A043G, and NNX14AQ89G and NSF grants AST-0808050, AST-1109911, and AST-1835400.
This research has made use of data from the MOJAVE database that is maintained by the MOJAVE team \citep{MOJAVE_XV}. The MOJAVE program is supported by NASA-{\it Fermi} grant 80NSSC19K1579.
This research has made use of the CIRADA cutout service at http://cutouts.cirada.ca, operated by the Canadian Initiative for Radio Astronomy Data Analysis (CIRADA). CIRADA is funded by a grant from the Canada Foundation for Innovation 2017 Innovation Fund (Project 35999), as well as by the Provinces of Ontario, British Columbia, Alberta, Manitoba and Quebec, in collaboration with the National Research Council of Canada, the US National Radio Astronomy Observatory and Australia’s Commonwealth Scientific and Industrial Research Organisation.
S.K. and K.T. acknowledge support from the European Research Council (ERC) under the European Unions Horizon 2020 research and innovation programme under grant agreement No.~771282.
KT acknowledges support from the Foundation of Research and Technology - Hellas Synergy Grants Program through project POLAR, jointly implemented by the Institute of Astrophysics and the Institute of Computer Science.
A.S. was supported by the NASA Contract NAS8-03060 to the Chandra X-ray Center.

This paper depended on a very large amount of \ac{vlbi} data, almost all of which was taken with the Very Long Baseline Array. The National Radio Astronomy Observatory is a facility of the National Science Foundation operated under cooperative agreement by Associated Universities, Inc.

\end{acknowledgments}

\appendix

\startlongtable
\begin{deluxetable*}{lllcccccccc}
\tablecaption{Structure Types and Literature References for the Combined PR, CJ1 and PW$^\dag$ Samples
\label{tab:samples}}
\tablehead{
    B1950 Name&J2000 Name&Alias&PR&CJ1&PW&Type&Optical&Large-scale&Small-scale\\
    &&&&&&&ID$^\ddag$&Structure&Structure\\
    (1)&(2)&(3)&(4)&(5)&(6)&(7)&(8)&(9)&(10)}
\startdata
 0010+775 	&	 J0013+7748 	&	          S5 0010+77&	   	&	 Y 	&	   	&	 CT,CSS  	&	  G 	&	     	&	 41,42  	\\
 0010+405 	&	 J0013+4051 	&	          4C +40.01 &	   	&	 Y 	&	   	&	 CT,CFS  	&	  G 	&	     	&	 41,42  	\\
 0013+790 	&	 J0016+7916 	&	            3C~6.1 	&	   	&	 Y 	&	 Y 	&	 FR~II  	&	  G 	&	   3,9,62 	&	        	\\
 0016+731 	&	 J0019+7327 	&	           S5 0016+73&	 Y 	&	   	&	 Y 	&	  U,CFS 	&	  Q 	&	9	&	31	\\
 0022+390 	&	 J0025+3919 	&	           S4 0022+39&	   	&	 Y 	&	   	&	 CT,CFS  	&	  Q 	&	     	&	 41,42  	\\
  0026+34 	&	 J0029+3456 	&	        B2 0026+34 	&	   	&	   	&	 Y 	&	 CSO 	&	  G 	&	    	&	    58, 67 	\\
  0038+32 	&	 J0040+3310 	&	             3C~19 	&	   	&	   	&	 Y 	&	 FR~II  	&	  G 	&	   8,9,62 	&	        	\\
 0040+517 	&	 J0043+5203 	&	             3C~20 	&	 Y 	&	   	&	 Y 	&	 FR~II  	&	  G 	&	   8,9,64 	&	        	\\
 0048+509 	&	 J0050+5112 	&	            3C 22.0&	   	&	 Y 	&	   	&	 FR~II  	&	  G 	&	8	&	        	\\
 0102+480 	&	 J0105+4819 	&	                   &	   	&	 Y 	&	   	&	 CT,CFS  	&	    	&	     	&	 41,42  	\\
  0104+32 	&	 J0107+3224 	&	             3C~31 	&	   	&	   	&	 Y$^\dag$ 	&	  FR~I 	&	  G 	&	   8,9 	&	        	\\
  0106+13 	&	 J0108+1320 	&	             3C~33 	&	   	&	   	&	 Y$^\dag$ 	&	  FR~II 	&	  G 	&	  9,34 	&	        	\\
 0106+729 	&	 J0109+7311 	&	           3C~33.1 	&	   	&	 Y 	&	 Y 	&	 FR~II  	&	  G 	&	   8,9 	&	        	\\
 0108+388 	&	 J0111+3906 	&	          S4 0108+388 	&	 Y 	&	   	&	   	&	 CSO 	&	  G 	&	    66 	&	    31, 41  	\\
  0116+31 	&	 J0119+3210 	&	        4C 31.04 	&	   	&	   	&	 Y$^\dag$ 	&	 CSO 	&	  G 	&	     	&	    35, 68  	\\
  0123+32 	&	 J0126+3313 	&	             3C~41 	&	   	&	   	&	 Y$^\dag$ 	&	 FR~II  	&	  G 	&	   5,9,62 	&	        	\\
  0125+28 	&	 J0128+2903 	&	             3C~42 	&	   	&	   	&	 Y 	&	 FR~II  	&	  G 	&	   8,9 	&	        	\\
  0127+23 	&	 J0129+2338 	&	             3C~43 	&	   	&	   	&	 Y 	&	 CSS  	&	  Q 	&	  9,48 	&	        	\\
  0133+20 	&	 J0136+2057 	&	             3C~47 	&	   	&	   	&	 Y 	&	 FR~II  	&	  Q 	&	   3,9 	&	        	\\
 0133+476 	&	 J0136+4751 	&	            OC 457 	&	 Y 	&	   	&	 Y 	&	 U,CFS  	&  Q 	&	9	&	31	\\
  0134+32 	&	 J0137+3309 	&	             3C~48 	&	   	&	   	&	 Y$^\dag$ 	&	  CSS 	&  Q 	&	  9,48 	&	        	\\
  0138+13 	&	 J0141+1353 	&	             3C~49 	&	   	&	   	&	 Y 	&	 CSS  	&  G 	&	  9,48 	&	        	\\
 0153+744 	&	 J0157+7442 	&	           S5 0153+744&	 Y 	&	   	&	 Y 	&	 U,CFS  	&	  Q 	&	9	&	31	\\
  0202+14 	&	 J0204+1514 	&	          4C~15.05 	&	   	&	   	&	 Y$^\dag$ 	&	 U,CFS 	&	 Q$^a$ 	&	9	&	58	\\
 0206+355 	&	 J0209+3547 	&	            4C +35.03&	   	&	 Y 	&	   	&	 FR~I  	&	  G 	&	3	&	        	\\
 0212+735 	&	 J0217+7349 	&	            S5 0212+73&	 Y 	&	   	&	 Y 	&	 U,CFS  	&	  Q 	&	9	&	31	\\
 0218+357 	&	 J0221+3556 	&	           B2 0218+357&	   	&	 Y 	&	   	&	  CT,CFS 	&	  Q 	&	     	&	 41,42  	\\
 0210+860 	&	 J0222+8619 	&	            3C~61.1 	&	 Y 	&	   	&	 Y 	&	 FR~II  	&	  G 	&	  9,34 	&	        	\\
 0220+427 	&	 J0223+4259 	&	            3C~66B 	&	 Y 	&	   	&	 Y 	&	 FR~I  	&	  G 	&	  9,46 	&	        	\\
 0220+397 	&	 J0223+4000 	&	             3C~65 	&	   	&	 Y 	&	 Y 	&	 FR~II  	&  G 	&	   5,9 	&	        	\\
  0221+27 	&	 J0224+2750 	&	             3C~67 	&	   	&	   	&	 Y 	&	 CSS  	&  G 	&	  9,48 	&	        	\\
  0223+34 	&	 J0226+3421 	&	          4C~34.07 	&	   	&	   	&	 Y$^\dag$ 	&	 CSS  	& Q$^a$ 	&	  9,57,56 	&	        	\\
  0235+16 	&	 J0238+1636 	&	           PKS 0235+164 	&	   	&	   	&	 Y$^\dag$ 	&	 U,CFS  	&	  Q 	&	9	&	58	\\
 0248+430 	&	 J0251+4315 	&	           B3 0248+430	&	   	&	 Y 	&	   	&	 CT,CFS  	&  Q 	&	     	&	 41,42  	\\
 0258+350 	&	 J0301+3512 	&	            NGC 1167&	   	&	 Y 	&	   	&	  FR~I 	&  G 	&	28	&	        	\\
  0300+16 	&	 J0303+1626 	&	           3C~76.1 	&	   	&	   	&	 Y$^\dag$ 	&	 FR~I  	&	  G 	&	  9,34 	&	        	\\
  0307+16 	&	 J0310+1705 	&	             3C~79 	&	   	&	   	&	 Y$^\dag$ 	&	 FR~II  	&	  G 	&	  9,23,62,64 	&	        	\\
 0307+444 	&	 J0310+4435 	&	            4C 44.07&	   	&	 Y 	&	   	&	   CSS	&	  Q 	&	53	&	        	\\
 0309+390 	&	 J0312+3916 	&	           4C 39.11&	   	&	 Y 	&	   	&	 CT,CSS  	&  G 	&	     	&	59	\\
 0314+416 	&	 J0318+4151 	&	          3C~83.1B 	&	 Y 	&	   	&	 Y 	&	 FR~I  	&	  G 	&	   6,9 	&	        	\\
  0316+16 	&	 J0318+1628 	&	          CTA 21&	   	&	   	&	 Y$^\dag$ 	&	 CSS  	&  G 	&	9	&	58	\\
 0316+413 	&	 J0319+4130 	&	             3C~84 	&	 Y 	&	   	&	 Y 	&	 U,CFS  	&	  G 	&	  9,33 	&	61,56	\\
  0319+12 	&	 J0321+1221 	&	           PKS 0319+12 	&	   	&	   	&	 Y 	&	 CSS  	& Q$^a$ 	&	9	&	54,56	\\
  0356+10 	&	 J0358+1026 	&	             3C~98 	&	   	&	   	&	 Y$^\dag$ 	&	 FR~II  	&	  G 	&	  9,47 	&	        	\\
  0400+25 	&	 J0403+2600 	&	           CTD 26 	&	   	&	   	&	 Y$^\dag$ 	&	 U,CFS  	&  Q 	&	9	&	58	\\
 0402+379 	&	 J0405+3803 	&	       4C +37.11  &	   	&	 Y 	&	   	&	 CSO 	&	  G 	&	     	&	 41,42  	\\
 0404+768 	&	 J0410+7656 	&	          4C~76.03 	&	 Y 	&	   	&	 Y 	&	  CSS 	&	  G 	&	  9,57 	&	        56	\\
 0407+747 	&	 J0413+7451 	&	          4C~74.08 	&	   	&	 Y 	&	 Y 	&	 FR~II  	&	 G? 	&	  9,19 	&	        	\\
  0410+11 	&	 J0413+1112 	&	            3C~109 	&	   	&	   	&	 Y$^\dag$ 	&	 FR~II  	&	  G 	&	  9,39 	&	        	\\
  0411+14 	&	 J0414+1416 	&	          4C~14.11 	&	   	&	   	&	 Y 	&	 FR~II  	&	  G 	&	  9,34,64 	&	        	\\
  0428+20 	&	 J0431+2037 	&	            PKS 0428+20 	&	   	&	   	&	 Y$^\dag$ 	&	  U,CFS 	&	 G$^a$	&	9	&	54,56	\\
  0433+29 	&	 J0437+2940 	&	            3C~123 	&	   	&	   	&	 Y$^\dag$ 	&	 FR~II  	&	  G 	&	  9,52,62,64 	&	        	\\
  0453+22 	&	 J0456+2249 	&	            3C~132 	&	   	&	   	&	 Y 	&	 FR~II  	&	  G 	&	   8,9,62,64	&	        	\\
 0454+844 	&	 J0508+8432 	&	             S5 0454+84&	 Y 	&	   	&	   	&	  CT,CFS 	&	  Q 	&	     	&	31	\\
  0518+16 	&	 J0521+1638 	&	            3C~138 	&	   	&	   	&	 Y$^\dag$ 	&	 CSS  	&	Q 	&	  9,48 	&	57	\\
  0528+13 	&	 J0530+1331 	&	            PKS 0528+134 	&	   	&	   	&	 Y$^\dag$ 	&	 U,CFS  	& Q$^a$ 	&	9	&	58	\\
 0538+498 	&	 J0542+4951 	&	            3C~147 	&	 Y 	&	   	&	 Y 	&	 CSS  	&  Q 	&	  9,48 	&	        	\\
 0602+673 	&	 J0607+6720 	&	                   	&	   	&	 Y 	&	   	&	 CT,CFS  	&	  Q 	&	     	&	 41,42  	\\
 0605+480 	&	 J0609+4804 	&	            3C~153 	&	 Y 	&	   	&	 Y 	&	 FR~II  	&	  G 	&	   3,9,62,64 	&	        	\\
 0620+389 	&	 J0624+3856 	&	            S4 0620+389	&	   	&	 Y 	&	   	&	  CT,CFS 	&	  Q 	&	     	&	 41,42  	\\
 0615+820 	&	 J0626+8202 	&	             S5 0615+82	&	   	&	 Y 	&	   	&	 CT,CFS  	&	  Q 	&	     	&	58	\\
 0642+449 	&	 J0646+4451 	&	             OH 471	&	   	&	 Y 	&	   	&	 CT,CSS  		&	  Q 	&	     	&	 41,42  	\\
 0646+600 	&	 J0650+6001 	&	             4 0646+60	&	   	&	 Y 	&	   	&	 CT,CFS  	&	  Q 	&	     	&	 41,42  	\\
 0650+371 	&	 J0653+3705 	&	             S4 0650+371 	&	   	&	 Y 	&	   	&	 CT,CFS  	&	  Q 	&	     	&	 41,42  	\\
 0651+542 	&	 J0655+5408 	&	            3C~171 	&	   	&	 Y 	&	 Y 	&	 FR~II  	&	  G 	&	  9,49,64 	&	        	\\
 0703+426 	&	 J0706+4230 	&	          4C~42.23 	&	   	&	 Y 	&	 Y 	&	 FR~I  	&	  G 	&	  9,19 	&	        	\\
 0702+749 	&	 J0709+7449 	&	            3C 173.1	&	   	&	 Y 	&	   	&	  FR~II 	&	  G 	&	34	&	        	\\
 0707+476 	&	 J0710+4732 	&	           S4 0707+47&	   	&	 Y 	&	   	&	 CT,CFS  	&	  Q 	&	     	&	 41,42  	\\
 0707+689 	&	 J0713+6852 	&	           4C 68.08	&	   	&	 Y 	&	   	&	  CSS 	&	  Q 	&	45	&	        	\\
 0710+439 	&	 J0713+4349 	&	       B3 0710+439 	&	 Y 	&	   	&	 Y 	&	 CSO 	&	  G 	&	     	&	    31, 69  	\\
 0711+356 	&	 J0714+3534 	&	            OI 318 	&	 Y 	&	   	&	 Y 	&	 U,CFS  	&		  Q 	&	9	&	31	\\
 0716+714 	&	 J0721+7120 	&	           TXS 0716+714 	&	   	&	 Y 	&	   	&	 CT,CFS  	&	  Q 	&	     	&	 41,42  	\\
 0723+679 	&	 J0728+6748 	&	            3C~179 	&	 Y 	&	   	&	 Y 	&	 FR~II  	&	  Q 	&	  9,19,62 	&	        	\\
  0735+17 	&	 J0738+1742 	&	            OI 158 	&	   	&	   	&	 Y$^\dag$ 	&	 U,CFS  	&	  Q 	&	9	&	58	\\
  0738+31 	&	 J0741+3112 	&	            OI 363 	&	   	&	   	&	 Y$^\dag$ 	&	  U,CFS 	&	  Q 	&	9	&	58	\\
 0734+805 	&	 J0743+8026 	&	          3C~184.1 	&	   	&	 Y 	&	 Y 	&	 FR~II  	&	  G 	&	  9,47 	&	        	\\
  0742+10 	&	 J0745+1011 	&	            OI 471 	&	   	&	   	&	 Y$^\dag$ 	&	  U,CFS 	&	 EF$^a$ 	&	9	&	58	\\
  0744+55 	&	 J0748+5548 	&	            DA~240 	&	   	&	   	&	 Y 	&	 FR~?  	&		  G 	&	  9,24 	&	        	\\
 0746+483 	&	 J0750+4814 	&	          S4 0746+483	&	   	&	 Y 	&	   	&	 CT,CFS  	&	  Q 	&	     	&	 41,42  	\\
  0748+12 	&	 J0750+1231 	&	            OI 280 	&	   	&	   	&	 Y$^\dag$ 	&	 U,CFS  	&	  Q 	&	9	&	58	\\
 0740+828 	&	 J0750+8241 	&	            S5 0740+82&	   	&	 Y 	&	   	&	 CT,CSS  	&	  Q 	&	     	&	 41,42  	\\
 0755+379 	&	 J0758+3747 	&	          NGC 2484	&	   	&	 Y 	&	 Y 	&	 CSS  	&	  G 	&	9	&	 41,42  	\\
  0802+24 	&	 J0805+2409 	&	            3C~192 	&	   	&	   	&	 Y$^\dag$ 	&	 FR~II  	&	  G 	&	  9,47 	&	        	\\
 0804+499 	&	 J0808+4950 	&	            OJ 508 	&	 Y 	&	   	&	 Y 	&	  U,CFS 	&	  Q 	&	9	&	31	\\
 0805+410 	&	 J0808+4052 	&	            S4 0805+41	&	   	&	 Y 	&	   	&	  CT,CFS 	&	  Q 	&	     	&	 41,42  	\\
 0809+483 	&	 J0813+4813 	&	            3C~196 	&	 Y 	&	   	&	 Y 	&	 FR~II  	&	  Q 	&	9.25	&	        	\\
 0812+367 	&	 J0815+3635 	&	            B2 0812+36 	&	   	&	 Y 	&	   	&	 CT,CFS  	&	  Q 	&	     	&	 41,42  	\\
 0814+425 	&	 J0818+4222 	&	            OJ 425 	&	 Y 	&	   	&	 Y 	&	 U,CFS  	&		  Q 	&	9	&	31	\\
 0816+526 	&	 J0819+5232 	&	           4C 52.18	&	   	&	 Y 	&	   	&	 FR~II  	&	  G 	&	13	&	        	\\
 0818+472 	&	 J0821+4702 	&	      3C~197.1             	&	   	&	 Y 	&	   	&	  FR~II 	&	  G 	&	   4,9,62 	&	        	\\
 0820+560 	&	 J0824+5552 	&	          OJ 535	&	   	&	 Y 	&	   	&	  CT,CFS 	&	  Q 	&	     	&	 41,42  	\\
 0821+394 	&	 J0824+3916 	&	        4C +39.23	&	   	&	 Y 	&	   	&	 CT,CFS  	&	  Q 	&	     	&	 41,42  	\\
 0827+378 	&	 J0831+3742 	&	        B2 0827+37	&	   	&	 Y 	&	   	&	  CT,CSS 	&	  Q 	&	     	&	 41,42  	\\
 0828+493 	&	 J0832+4913 	&	        OJ 448 	&	   	&	 Y 	&	   	&	 CT,CFS  	&	  Q 	&	     	&	 41,42  	\\
 0831+557 	&	 J0834+5534 	&	          4C~55.16 	&	 Y 	&	   	&	 Y 	&	 U,CFS  	&	  G 	&	9	&	31	\\
 0833+585 	&	 J0837+5825 	&	           S4 0833+58 	&	   	&	 Y 	&	   	&	  CT,CFS 	&	  Q 	&	     	&	 41,42  	\\
  0838+13 	&	 J0840+1312 	&	            3C~207 	&	   	&	   	&	 Y$^\dag$ 	&	 FR~II  	&	  Q 	&	   3,9,62 	&	        	\\
 0836+710 	&	 J0841+7053 	&	          4C~71.07 	&	 Y 	&	   	&	 Y 	&	 U,CFS  	&	 Q$^a$ 	&	9	&	31	\\
 0844+540 	&	 J0847+5352 	&	          NGC 2656	&	   	&	 Y 	&	   	&	  FR~I 	&	  G 	&	12	&	        	\\
 0850+581 	&	 J0854+5757 	&	          4C~58.17 	&	 Y 	&	   	&	   	&	  CT,CFS 		&	  Q 	&	     	&	31	\\
  0851+20 	&	 J0854+2006 	&	            OJ 287 	&	   	&	   	&	 Y$^\dag$ 	&	  U,CFS 		&	  Q 	&	9	&	58	\\
 0859+470 	&	 J0903+4651 	&	          4C~47.29 	&	 Y 	&	   	&	 Y 	&	 U,CFS  	&		  Q 	&	9	&	  n,31  	\\
 0900+428 	&	 J0904+4238 	&	          B3 0900+428 	&	   	&	 Y 	&	   	&	 CT,CFS  		&	  G 	&	     	&	 41,42  	\\
 0906+430 	&	 J0909+4253 	&	            3C~216 	&	 Y 	&	   	&	 Y 	&	 U,CFS  	&	  Q 	&	  9,48,62	&	31	\\
 0917+449 	&	 J0920+4441 	&	           TXS 0917+449 	&	   	&	 Y 	&	   	&	 CT,CFS  	&	  Q 	&	     	&	 41,42  	\\
 0917+458 	&	 J0921+4538 	&	            3C~219 	&	 Y 	&	   	&	 Y 	&	 FR~II  	&	  G 	&	  9,36 	&	        	\\
 0917+624 	&	 J0921+6215 	&	           OK 630	&	   	&	 Y 	&	   	&	  CT,CFS 	&	  Q 	&	     	&	 41,42  	\\
 0923+392 	&	 J0927+3902 	&	          4C~39.25 	&	 Y 	&	   	&	 Y 	&	 U,CFS  	&	  Q 	&	9	&	60	\\
 0936+361 	&	 J0939+3553 	&	            3C~223 	&	   	&	 Y 	&	 Y 	&	 FR~II  	&	  G 	&	  9,47 	&	        	\\
 0938+399 	&	 J0941+3944 	&	           3C 223.1 	&	   	&	 Y 	&	   	&	 Complex  	&	  G 	&	50	&	        	\\
  0939+14 	&	 J0942+1345 	&	           3C~225B 	&	   	&	   	&	 Y 	&	 FR~II  	&	  G 	&	   8,9 	&	        	\\
 0945+408 	&	 J0948+4039 	&	          4C~40.24 	&	 Y 	&	   	&	   	&	  CT,CFS 	&		  Q 	&	     	&	31	\\
 0945+664 	&	 J0949+6615 	&	          4C~66.09 	&	   	&	 Y 	&	 Y 	&	 U,CFS  	&	  G 	&	9	&	 41,42  	\\
  0945+73 	&	 J0949+7314 	&	4C 73.08                   	&	   	&	   	&	 Y 	&	 FR~II  	&	  G 	&	7	&	        	\\
  0947+14 	&	 J0950+1420 	&	            3C~228 	&	   	&	   	&	 Y 	&	 FR~II  	&	  G 	&	   8,9,62 	&	        	\\
 0951+699 	&	 J0955+6940 	&	      M82$^*$ (3C~231) 	&	 Y 	&	   	&	 Y 	&	 FR~?  	&	  G 	&	  9,10 	&	        	\\
 0954+556 	&	 J0957+5522 	&	          4C~55.17 	&	 Y 	&	   	&	 Y 	&	 U,CFS  	&	  Q 	&	9	&	31	\\
 0955+476 	&	 J0958+4725 	&	           OK 492	&	   	&	 Y 	&	   	&	 CT,CFS  	&	  Q 	&	     	&	 41,42  	\\
 0954+658 	&	 J0958+6533 	&	         S4 0954+65 	&	 Y 	&	   	&	   	&	  CT,CFS 	&	  Q 	&	     	&	58	\\
  0958+29 	&	 J1001+2847 	&	            3C~234 	&	   	&	   	&	 Y$^\dag$ 	&	 FR~II  	&	  G 	&	  9,29,64 	&	        	\\
 1003+351 	&	 J1006+3454 	&	            3C~236 	&	 Y 	&	   	&	 Y 	&	 FR~II  	&	  G 	&	  9,24 	&	        	\\
 1003+830 	&	 J1010+8250 	&	          S5 1003+83 	&	   	&	 Y 	&	   	&	 CT,CFS  	&	  G 	&	     	&	 41,42  	\\
 1007+416 	&	 J1010+4132 	&	         4C 41.21 	&	   	&	 Y 	&	   	&	  FR~II 	&	  Q 	&	22	&	22	\\
 1015+359 	&	 J1018+3542 	&	           B2 1015+35B	&	   	&	 Y 	&	   	&	 CT,CFS  	&	  Q 	&	     	&	 41,42  	\\
 1020+400 	&	 J1023+3948 	&	          4C +40.25&	   	&	 Y 	&	   	&	 CT,CFS  	&	  Q 	&	     	&	 41,42  	\\
 1030+415 	&	 J1033+4116 	&	         IVS B1030+415	&	   	&	 Y 	&	   	&	  CT ,CSF	&	  Q 	&	     	&	 41,42  	\\
 1030+585 	&	 J1033+5814 	&	          3C~244.1 	&	   	&	 Y 	&	 Y 	&	 FR~II  	&	  G 	&	   8,9 	&	        	\\
 1031+567 	&	 J1035+5628 	&	   JVAS~J1035+5628 	&	 Y 	&	   	&	 Y 	&	 CSO 	&	 G$^a$ 	&	     	&	    65, 67  	\\
  1040+12 	&	 J1042+1203 	&	            3C~245 	&	   	&	   	&	 Y$^\dag$ 	&	  D 	&	  Q 	&	   8,9 	&	        	\\
 1039+811 	&	 J1044+8054 	&	          S5 1039+81 	&	   	&	 Y 	&	   	&	 CT,CFS  	&	  Q 	&	     	&	42	\\
 1044+719 	&	 J1048+7143 	&	          S5 1044+71&	   	&	 Y 	&	   	&	 CT,CFS  	&	  Q 	&	     	&	 41,42  	\\
 1053+704 	&	 J1056+7011 	&	         S5 1053+70	&	   	&	 Y 	&	   	&	  CT,CFS 	&	  Q 	&	     	&	 41,42  	\\
 1053+815 	&	 J1058+8114 	&	         S5 1053+81 	&	   	&	 Y 	&	   	&	  CT,CFS 	&	  G 	&	     	&	 41,42  	\\
  1055+20 	&	 J1058+1951 	&	          4C~20.24 	&	   	&	   	&	 Y 	&	 D  	&	  Q 	&	9	&	58	\\
 1056+432 	&	 J1058+4301 	&	            3C~247 	&	   	&	 Y 	&	 Y 	&	  FR~II 	&	  G 	&	  9,40,62 	&	        	\\
 1058+726 	&	 J1101+7225 	&	           S5 1058+726 	&	   	&	 Y 	&	   	&	 CT,CSS  	&	  Q 	&	     	&	 41,42  	\\
 1100+772 	&	 J1104+7658 	&	         3C 249.1	&	   	&	 Y 	&	   	&	  FR~II 	&	  Q 	&	37	&	        	\\
 1101+384 	&	 J1104+3812 	&	        Mrk 421  	&	   	&	 Y 	&	   	&	 CT,CFS  	&	  Q 	&	     	&	 41,42  	\\
 1111+408 	&	 J1114+4037 	&	         3C 254   	&	   	&	 Y 	&	   	&	  FR~II 	&	  Q 	&	55	&	        	\\
  1116+12 	&	 J1118+1234 	&	          4C~12.39 	&	   	&	   	&	 Y$^\dag$ 	&	 U,CFS   &	  Q 	&	9	&	21	\\
 1128+385 	&	 J1130+3815 	&	          IVS B1128+385 	&	   	&	 Y 	&	   	&	 CT,CFS  	&	  Q 	&	     	&	 41,42  	\\
 1137+660 	&	 J1139+6547 	&	            3C~263 	&	   	&	 Y 	&	 Y 	&	 FR~II  	&		  Q 	&	  9,37 	&	        	\\
 1138+594 	&	 J1140+5912 	&	            4C +59.16 	&	   	&	 Y 	&	   	&	  CT,CSS 	&	    	&	     	&	 41,42  	\\
  1140+22 	&	 J1143+2206 	&	          3C~263.1 	&	   	&	   	&	 Y 	&	 FR~II  	&	  G 	&	   8,9 	&	        	\\
  1142+19 	&	 J1145+1936 	&	            3C~264 	&	   	&	   	&	 Y$^\dag$ 	&	 FR~I  	&		  G 	&	  9,26 	&	        	\\
 1144+542 	&	 J1146+5356 	&	         S4 1144+54	&	   	&	 Y 	&	   	&	 CT,CSS  	&	  Q 	&	     	&	 41,42  	\\
 1144+402 	&	 J1146+3958 	&	         S4 1144+40	&	   	&	 Y 	&	   	&	  CT,CFS 	&	  Q 	&	     	&	 41,42  	\\
 1150+812 	&	 J1153+8058 	&	         S5 1150+81 	&	   	&	 Y 	&	   	&	 CT,CFS  	&	  Q 	&	     	&	58	\\
 1150+497 	&	 J1153+4931 	&	          4C~49.22 	&	   	&	 Y 	&	 Y 	&	 CSS  	&  Q 	&	9	&	 41,42  	\\
 1152+551 	&	 J1155+5453 	&	          4C +55.22 	&	   	&	 Y 	&	   	&	  Complex 	&	  G 	&	16	&	        	\\
  1153+31 	&	 J1156+3128 	&	          4C~31.38 	&	   	&	   	&	 Y 	&	 CSS  	&  Q 	&	  9,57 	&	        	\\
 1157+732 	&	 J1200+7300 	&	          3C~268.1 	&	 Y 	&	   	&	 Y 	&	 FR~II  	&	   G$^a$ 	&	   8,9,62 	&	        	\\
 1203+645 	&	 J1206+6413 	&	          3C~268.3 	&	   	&	 Y 	&	 Y 	&	 CSS  	&	  G 	&	  9,48 	&	        	\\
 1213+538 	&	 J1215+5335 	&	          4C +53.24	&	   	&	 Y 	&	   	&	  FR~II 	&	  Q 	&	44	&	        	\\
 1213+350 	&	 J1215+3448 	&	           S4 1213+350&	   	&	 Y 	&	   	&	 CT,CFS  	&	  Q 	&	     	&	 41,42  	\\
 1216+487 	&	 J1219+4829 	&	           S4 1216+48	&	   	&	 Y 	&	   	&	 CT,CFS  	&	  Q 	&	     	&	 41,42  	\\
  1218+33 	&	 J1220+3343 	&	          3C~270.1 	&	   	&	   	&	 Y 	&	 FR~II  	&	  Q 	&	   8,9,62 	&	        	\\
  1222+13 	&	 J1225+1253 	&	          3C~272.1 	&	   	&	   	&	 Y$^\dag$ 	&	 FR~I  	&	  G 	&	   8,9	&	        	\\
 1225+368 	&	 J1227+3635 	&	         B2 1225+36 	&	   	&	 Y 	&	 Y 	&	 CSO 	&	 Q$^b$	&	     	&	    54, 56, 70  	\\
  1228+12 	&	 J1230+1223 	&	            3C~274 	&	   	&	   	&	 Y$^\dag$ 	&	 FR~I  	&	  G 	&	  1,9  	&	        	\\
  1241+16 	&	 J1243+1622 	&	          3C~275.1 	&	   	&	   	&	 Y 	&	 FR~II  	&	  Q 	&	   8,9,62 	&	        	\\
 1242+410 	&	 J1244+4048 	&	       B3 1242+410 	&	   	&	 Y 	&	   	&	 CSO 	&	  Q 	&	     	&	 41,42  	\\
 1250+568 	&	 J1252+5634 	&	          3C~277.1 	&	   	&	 Y 	&	 Y 	&	 CSS  	&	  Q 	&	  9,43 	&	        	\\
  1251+27 	&	 J1254+2737 	&	          3C~277.3 	&	   	&	   	&	 Y 	&	 FR~II  	&	  G 	&	   3,9 	&	        	\\
 1254+476 	&	 J1256+4720 	&	            3C~280 	&	 Y 	&	   	&	 Y 	&	  FR~II 	&	  G 	&	  9,17 	&	        	\\
 1311+678 	&	 J1313+6736 	&	           4C +67.22 	&	   	&	 Y 	&	   	&	 CSS  	&	    	&	9	&	 41,42  	\\
 1317+520 	&	 J1319+5148 	&	           4C +52.27	&	   	&	 Y 	&	   	&	 CT,CFS  	&	  Q 	&	     	&	 41,42  	\\
 1319+428 	&	 J1321+4235 	&	           3C 285 	&	   	&	 Y 	&	   	&	  FR~II 	&	  G 	&	50	&	        	\\
  1323+32 	&	 J1326+3154 	&	            DA~344 	&	   	&	   	&	 Y$^\dag$ 	&	 CSO 	&	 G$^a$	&	     	&	    58, 56, 70  	\\
  1328+25 	&	 J1330+2509 	&	            3C~287 	&	   	&	   	&	 Y$^\dag$ 	&	 CSS  	&	  Q 	&	9	&	    58  	\\
  1328+30 	&	 J1331+3030 	&	            3C~286 	&	   	&	   	&	 Y$^\dag$ 	&	 CSS  	&	  Q 	&	  9,48 	&	        	\\
 1333+459 	&	 J1335+4542 	&	         S4 1333+459 	&	   	&	 Y 	&	   	&	 CT,CFS  	&	  Q 	&	     	&	 41,42  	\\
 1333+589 	&	 J1335+5844 	&	   JVAS~J1335+5844 	&	   	&	 Y 	&	   	&	 CSO 	&	    	&	     	&	 41,42  	\\
 1336+391 	&	 J1338+3851 	&	            3C~288 	&	   	&	 Y 	&	 Y 	&	 FR~?  	&	  G 	&	  9,32 	&	        	\\
 1342+663 	&	 J1344+6606 	&	        S4 1342+663 	&	   	&	 Y 	&	   	&	 CT,CFS  	&	  Q 	&	     	&	 41,42  	\\
  1345+12 	&	 J1347+1217 	&	     PKS~B1345+125 	&	   	&	   	&	 Y$^\dag$ 	&	 CSO 	&	  G 	&	  71   	&	    58  	\\
 1347+539 	&	 J1349+5341 	&	       4C 53.28	&	   	&	 Y 	&	   	&	 CT,CFS  	&	  Q 	&	     	&	 41,42,72  	\\
 1349+647 	&	 J1350+6429 	&	        3C 292  	&	   	&	 Y 	&	   	&	 FR~II  	&	  G 	&	30	&	        	\\
  1350+31 	&	 J1352+3126 	&	            3C~293 	&	   	&	   	&	 Y$^\dag$ 	&	 FR~?  	&	  G 	&	  9,43 	&	        	\\
  1354+19 	&	 J1357+1919 	&	          4C~19.44 	&	   	&	   	&	 Y$^\dag$ 	&	 FR~II  	&	  Q 	&	9	&	58	\\
 1357+769 	&	 J1357+7643 	&	         S5 1357+76 	&	   	&	 Y 	&	   	&	 CT,CFS  		&	  Q 	&	     	&	 41,42  	\\
 1358+624 	&	 J1400+6210 	&	         4C 62.22 	&	 Y 	&	   	&	 Y 	&	 CSO 		&	  G 	&	     	&	    56,70  	\\
  1404+28 	&	 J1407+2827 	&	           OQ 208 	&	   	&	   	&	 Y$^\dag$ 	&	 CSO 	&	  G 	&	     	&	    58, 72  	\\
 1409+524 	&	 J1411+5212 	&	            3C~295 	&	 Y 	&	   	&	 Y 	&	 FR~II  		&	  G 	&	  9,17 	&	        	\\
  1413+34 	&	 J1416+3444 	&	        B2 1413+34 	&	   	&	   	&	 Y 	&	 CSO 	&  EF$^a$	&	   	&	      54,56  	\\
  1414+11 	&	 J1416+1048 	&	            3C~296 	&	   	&	   	&	 Y$^\dag$ 	&	 FR~I  	&	  G 	&	  9,34 	&	        	\\
 1418+546 	&	 J1419+5423 	&	            OQ 530	&	   	&	 Y 	&	   	&	 CT,CFS  	&	  Q 	&	     	&	 41,42  	\\
 1419+419 	&	 J1421+4144 	&	            3C~299 	&	   	&	 Y 	&	 Y 	&	 CSS  	&	  G 	&	  9,43 	&	        	\\
  1420+19 	&	 J1422+1935 	&	            3C~300 	&	   	&	   	&	 Y 	&	 FR~II  		&	  G 	&	  9,29,64 	&	        	\\
 1435+638 	&	 J1436+6336 	&	          VIPS 0792	&	   	&	 Y 	&	   	&	 CT,CFS  	&	  Q 	&	     	&	 41,42  	\\
 1437+624 	&	 J1438+6211 	&	          OQ 663	&	   	&	 Y 	&	   	&	 CT,CSS  		&	  Q 	&	     	&	 41,42  	\\
 1438+385 	&	 J1440+3820 	&	          S4 1438+38	&	   	&	 Y 	&	   	&	 CT,CFS  		&	  Q 	&	     	&	 41,42  	\\
 1441+522 	&	 J1443+5201 	&	            3C~303 	&	   	&	 Y 	&	 Y 	&	 FR~II  	&	  G 	&	  9,17 	&	        	\\
  1442+10 	&	 J1445+0958 	&	            OQ 172 	&	   	&	   	&	 Y 	&	 CSS  	&	  Q 	&	9	&	 51,56  	\\
 1448+634 	&	 J1449+6316 	&	            3C~305 	&	   	&	 Y 	&	 Y 	&	 FR~? 	&	  G 	&	  9,17 	&	        	\\
 1458+718 	&	 J1459+7140 	&	          3C~309.1 	&	 Y 	&	   	&	 Y 	&	 CSS  	&	  Q 	&	  9,48 	&	31	\\
  1502+10 	&	 J1504+1029 	&	          4C~10.39 	&	   	&	   	&	 Y$^\dag$ 	&	 U,CFS  		&	  Q 	&	9	&	58	\\
  1502+26 	&	 J1504+2600 	&	            3C~310 	&	   	&	   	&	 Y$^\dag$ 	&	 FR~II  	&	  G 	&	27	&	        	\\
 1504+377 	&	 J1506+3730 	&	           B2 1504+37 	&	   	&	 Y 	&	   	&	 CT,CFS  	&	  G 	&	     	&	 41,42  	\\
  1511+26 	&	 J1513+2607 	&	            3C~315 	&	   	&	   	&	 Y 	&	 FR~?  	&	  G 	&	  9,29 	&	        	\\
  1529+24 	&	 J1531+2404 	&	            3C~321 	&	   	&	   	&	 Y 	&	 FR~II  	&	  G 	&	   8,9 	&	        	\\
  1538+14 	&	 J1540+1447 	&	          4C~14.60 	&	   	&	   	&	 Y$^\dag$ 	&	 U,CFS  	&	 Q? 	&	9	&	58	\\
 1547+507 	&	 J1549+5038 	&	           S4 1547+507 	&	   	&	 Y 	&	   	&	 CT,CFS  	&	  Q 	&	     	&	 41,42  	\\
 1549+628 	&	 J1549+6241 	&	            3C~325 	&	   	&	 Y 	&	 Y 	&	 FR~II  	&	  G 	&	   8,9 	&	        	\\
 1557+708 	&	 J1557+7041 	&	          NGC 6048 	&	   	&	 Y 	&	 Y 	&	 FR~I  	&	  G 	&	  9,19 	&	        	\\
  1600+33 	&	 J1602+3326 	&	          4C +33.38 	&	   	&	   	&	 Y$^\dag$ 	&	 CSS  	&	   G$^a$ 	&	9	&	54,56	\\
  1607+26 	&	 J1609+2641 	&	            CTD 93 	&	   	&	   	&	 Y$^\dag$ 	&	 CSO 	&	 G$^a$ 	&	     	&	    58,73 	\\
 1609+660 	&	 J1609+6556 	&	            3C~330 	&	 Y 	&	   	&	 Y 	&	 FR~II  	&	  G 	&	8	&	        	\\
  1611+34 	&	 J1613+3412 	&	           DA 406 	&	   	&	   	&	 Y$^\dag$ 	&	  CT,CFS 	&	  Q 	&	     	&	58	\\
 1624+416 	&	 J1625+4134 	&	          4C~41.32 	&	 Y 	&	   	&	 Y 	&	 U,CFS 	&	 Q$^a$ 	&	9	&	31	\\
 1627+444 	&	 J1628+4419 	&	            3C~337 	&	   	&	 Y 	&	 Y 	&	 FR~II  	&		  G 	&	   8,9 	&	        	\\
 1637+826 	&	 J1632+8232 	&	          NGC~6251 	&	   	&	 Y 	&	 Y 	&	 FR~?  	&	  G 	&	  11 	&	        	\\
 1634+628 	&	 J1634+6245 	&	            3C~343 	&	 Y 	&	   	&	 Y 	&	 CSS  	&	  Q 	&	  9,57 	&	        	\\
 1633+382 	&	 J1635+3808 	&	          4C~38.41 	&	 Y 	&	   	&	 Y 	&	 U,CFS  	&		  Q 	&	     	&	31	\\
 1637+574 	&	 J1638+5720 	&	            OS 562 	&	 Y 	&	   	&	   	&	  CT,CFS 	&	  Q 	&	     	&	31	\\
 1637+626 	&	 J1638+6234 	&	          3C~343.1 	&	   	&	 Y 	&	 Y 	&	  CSS 	&	  G 	&	  9,57 	&	        	\\
 1638+398 	&	 J1640+3946 	&	          NRAO 512 	&	   	&	 Y 	&	   	&	 CT,CFS  	&	  Q 	&	     	&	 41,42  	\\
 1642+690 	&	 J1642+6856 	&	          4C~69.21 	&	 Y 	&	   	&	   	&	 CT,CFS  	&	  Q 	&	     	&	31	\\
 1641+399 	&	 J1642+3948 	&	            3C~345 	&	 Y 	&	   	&	 Y 	&	 U,CFS  	&	  Q 	&	9	&	58	\\
  1641+17 	&	 J1643+1715 	&	            3C~346 	&	   	&	   	&	 Y$^\dag$ 	&	 FR~I  	&	  G 	&	   3,9 	&	        	\\
 1652+398 	&	 J1653+3945 	&	          Mrk 501 	&	 Y 	&	   	&	   	&	 CT,CFS  	&	  G 	&	     	&	31	\\
 1656+482 	&	 J1657+4808 	&	          4C +48.41 	&	   	&	 Y 	&	   	&	 CT,CFS  	&	    	&	     	&	 41,42  	\\
 1656+477 	&	 J1658+4737 	&	          S4 1656+47 	&	   	&	 Y 	&	   	&	  CT,CFS 	&	  Q 	&	     	&	 41,42  	\\
 1658+471 	&	 J1659+4702 	&	            3C~349 	&	   	&	 Y 	&	 Y 	&	 FR~II  	&		  G 	&	   8,9,64 	&	        	\\
 1704+608 	&	 J1704+6044 	&	            3C~351 	&	   	&	 Y 	&	 Y 	&	 FR~II  	&		  Q 	&	  4,9 	&	        	\\
 1719+357 	&	 J1721+3542 	&	         S4 1719+357 	&	   	&	 Y 	&	   	&	 CT,CFS  	&		  Q 	&	     	&	 41,42  	\\
  1726+31 	&	 J1728+3145 	&	            3C~357 	&	   	&	   	&	 Y 	&	 FR~II  	&		  G 	&	  1,9  	&	        	\\
 1732+389 	&	 J1734+3857 	&	          OT 355 	&	   	&	 Y 	&	   	&	 CT,CFS  	&	  Q 	&	     	&	 41,42  	\\
 1734+508 	&	 J1735+5049 	&	  	&	   	&	 Y 	&	   	&	 CSO 	&		    	&	     	&	 41,42,74  	\\
 1738+476 	&	 J1739+4737 	&	          OT 465  	&	   	&	 Y 	&	   	&	  CT,CFS 	&	  Q 	&	     	&	 41,42  	\\
 1739+522 	&	 J1740+5211 	&	          4C~51.37 	&	 Y 	&	   	&	 Y 	&	 U,CFS  	&		  Q 	&	9	&	31	\\
 1749+701 	&	 J1748+7005 	&	        S4 1749+70 	&	 Y 	&	   	&	 Y 	&	 CSS  	&	  Q 	&	9	&	31	\\
 1751+441 	&	 J1753+4409 	&	        S4 1751+441 	&	   	&	 Y 	&	   	&	 CT,CFS  	&	  Q 	&	     	&	 41,42  	\\
 1758+388 	&	 J1800+3848 	&	        B3 1758+388B	&	   	&	 Y 	&	   	&	 CT,CFS  	&	  Q 	&	     	&	 41,42  	\\
 1803+784 	&	 J1800+7828 	&	         S5 1803+784 	&	 Y 	&	   	&	 Y 	&	 U,CFS  	&	  Q 	&	9	&	31	\\
 1800+440 	&	 J1801+4404 	&	         S4 1800+44	&	   	&	 Y 	&	   	&	 CT,CFS  	&	  Q 	&	     	&	 41,42  	\\
 1807+698 	&	 J1806+6949 	&	            3C~371 	&	 Y 	&	   	&	 Y 	&	 U,CFS  	&	  G 	&	9	&	31	\\
 1819+396 	&	 J1821+3942 	&	          4C~39.56 	&	   	&	 Y 	&	 Y 	&	 CSS  	&	 G? 	&	  9,57,56 	&	        	\\
 1823+568 	&	 J1824+5651 	&	          4C~56.27 	&	 Y 	&	   	&	 Y 	&	 U,CFS  	&	  Q 	&	9	&	31	\\
 1825+743 	&	 J1824+7420 	&	3C 379.1                   	&	   	&	 Y 	&	   	&	 FR~II  	&	  G 	&	1	&	        	\\
 1828+487 	&	 J1829+4844 	&	            3C~380 	&	 Y 	&	   	&	 Y 	&	 D  	&	  Q 	&	  9,48 	&	        	\\
  1829+29 	&	 J1831+2907 	&	          4C~29.56 	&	   	&	   	&	 Y 	&	 CSS  	&	 G$^a$ 	&	  9,57,56 	&	        	\\
 1833+653 	&	 J1833+6521 	&	   3C~383          	&	   	&	 Y 	&	   	&	 FR~II  	&	  G 	&	12	&	        	\\
 1832+474 	&	 J1833+4727 	&	            3C~381 	&	   	&	 Y 	&	 Y 	&	 FR~II  	&	  G 	&	  9,17,64 	&	        	\\
 1845+797 	&	 J1842+7946 	&	          3C~390.3 	&	 Y 	&	   	&	 Y 	&	 FR~II  	&		  G 	&	  9,17 	&	31	\\
 1842+681 	&	 J1842+6809 	&	          TXS 1842+681 	&	   	&	 Y 	&	   	&	 CT,CFS  	&	  Q 	&	     	&	 41,42  	\\
 1842+455 	&	 J1844+4533 	&	            3C~388 	&	 Y 	&	   	&	 Y 	&	 FR~II  	&	  G 	&	  9,17 	&	        	\\
 1843+356 	&	 J1845+3541 	&	    COINS J1845+3541 	&	   	&	 Y 	&	   	&	 CT,CFS  	&	  G 	&	     	&	 41,42  	\\
 1926+611 	&	 J1927+6117 	&	    S4 1926+61	&	   	&	 Y 	&	   	&	 CT,CFS  	&	  Q 	&	     	&	 41,42  	\\
 1928+738 	&	 J1927+7358 	&	          4C~73.18 	&	 Y 	&	   	&	 Y 	&	 U,CFS  	&	  Q 	&	9	&	31	\\
 1939+605 	&	 J1940+6041 	&	            3C~401 	&	 Y 	&	   	&	 Y 	&	 FR~II  	&	  G 	&	   8,9,64 	&	        	\\
 1940+504 	&	 J1941+5035 	&	              3C~402     	&	   	&	 Y 	&	   	&	FR~?  	&		  G 	&	4	&	        	\\
 1943+546 	&	 J1944+5448 	&	     COINS J1944+5448 	&	   	&	 Y 	&	   	&	 CSO 	&	  G 	&	     	&	 41,42  	\\
 1954+513 	&	 J1955+5131 	&	            OV 591 	&	 Y 	&	   	&	 Y 	&	 U,CFS  	&	  Q 	&	9	&	31	\\
 2007+777 	&	 J2005+7752 	&	        S5 2007+77  	&	   	&	 Y 	&	   	&	 CT,CFS  	&	  Q 	&	     	&	58	\\
 2010+723 	&	 J2009+7229 	&	         4C +72.28 	&	   	&	 Y 	&	   	&	 CT,CFS  	&	  Q 	&	     	&	 41,42  	\\
 2021+614 	&	 J2022+6136 	&	      TXS 2021+614 	&	 Y 	&	   	&	 Y 	&	 CSO 	&	  G 	&	     	&	    31,41  	\\
 2104+763 	&	 J2104+7633 	&	          3C~427.1 	&	   	&	 Y 	&	 Y 	&	 FR~II  	&	  G 	&	  9,17 	&	        	\\
  2121+24 	&	 J2123+2504 	&	            3C~433 	&	   	&	   	&	 Y$^\dag$ 	&	 FR~?  	&	  G 	&	   3,9 	&	        	\\
  2141+27 	&	 J2144+2810 	&	            3C~436 	&	   	&	   	&	 Y 	&	 FR~II  	&	  G 	&	   4,9,64 	&	        	\\
  2145+15 	&	 J2147+1520 	&	            3C~437 	&	   	&	   	&	 Y 	&	 FR~II  	&	  G 	&	   5,9 	&	        	\\
 2153+377 	&	 J2155+3800 	&	            3C~438 	&	 Y 	&	   	&	 Y 	&	 FR~II  	&		  G 	&	  9,17,64 	&	        	\\
 2200+420 	&	 J2202+4216 	&	            BL Lac 	&	 Y 	&	   	&	 Y 	&	 U,CFS  	&	  Q 	&	9	&	  58     	\\
  2203+29 	&	 J2206+2929 	&	            3C~441 	&	   	&	   	&	 Y 	&	 FR~II  	&	  G 	&	   5,9 	&	        	\\
 2207+374 	&	 J2209+3742 	&	           S4 2207+37 	&	   	&	 Y 	&	   	&	 CT,CSS  	&	  Q 	&	     	&	 41,42  	\\
 2214+350 	&	 J2216+3518 	&	            OY 324	&	   	&	 Y 	&	   	&	 CT,CFS  	&	  Q 	&	     	&	 41,42  	\\
 2229+695 	&	 J2230+6946 	&	          S5 2229+69  	&	   	&	 Y 	&	   	&	 CT,CFS  	&	  G 	&	     	&	 41,42  	\\
 2229+391 	&	 J2231+3921 	&	            3C~449 	&	 Y 	&	   	&	 Y 	&	 FR~1  	&	  G 	&	  9,18 	&	        	\\
  2230+11 	&	 J2232+1143 	&	          CTA 102 	&	   	&	   	&	 Y$^\dag$ 	&	  U,CFS 	&	  Q 	&	9	&	58,56	\\
 2243+394 	&	 J2245+3941 	&	            3C~452 	&	 Y 	&	   	&	 Y 	&	 FR~II  	&	  G 	&	  1,9  	&	        	\\
  2247+14 	&	 J2250+1419 	&	          4C~14.82 	&	   	&	   	&	 Y 	&	  CSS 	&  Q 	&	9	&	38	\\
  2251+15 	&	 J2253+1608 	&	          3C~454.3 	&	   	&	   	&	 Y$^\dag$ 	&	 U,CFS  	&	  Q 	&	  9,58 	&	        	\\
  2252+12 	&	 J2255+1313 	&	            3C~455 	&	   	&	   	&	 Y 	&	 FR~II  	&	  Q 	&	9	&	63	\\
 2253+417 	&	 J2255+4202 	&	          OY 489 	&	   	&	 Y 	&	   	&	 CT,CFS  	&	  Q 	&	     	&	 41,42  	\\
 2255+416 	&	 J2257+4154 	&	          4C 41.45 	&	   	&	 Y 	&	   	&	 CT,CSS  	&	  Q 	&	     	&	 41,42  	\\
 2311+469 	&	 J2313+4712 	&	           4C 46.47	&	   	&	 Y 	&	   	&	 CT,CSS  	&	  Q 	&	     	&	 41,42  	\\
 2323+435 	&	 J2325+4346 	&	          S4 2323+43 	&	   	&	 Y 	&	   	&	  CSS 	&  G 	&	45	&	        	\\
 2324+405 	&	 J2326+4048 	&	            3C~462 	&	   	&	 Y 	&	 Y 	&	 FR~II  	&	 G$^a$ 	&	9	&	 19  	\\
  2335+26 	&	 J2338+2701 	&	            3C~465 	&	   	&	   	&	 Y$^\dag$ 	&	 FR~I  	&	  G 	&	  1,9  	&	        	\\
 2342+821 	&	 J2344+8226 	&	          S5 2342+82 	&	 Y 	&	   	&	 Y 	&	  CSS 	&	    Q$^a$	&	  9,57,56 	&	        	\\
 2351+456 	&	 J2354+4553 	&	          4C~45.51 	&	 Y 	&	   	&	 Y 	&	 U,CFS  	&	 Q$^a$ 	&	9	&	31	\\
 2352+495 	&	 J2355+4950 	&	      DA 611 	&	 Y 	&	   	&	 Y 	&	 CSO 	&	  G 	&	     	&	    31,75	\\
\enddata
\tablecomments{$^*$ M82 s not an active galaxy, and was therefore not included in the statistical analyses in this paper. $^\dag$ indicates objects in the PWS subsample (see text). $^\ddag$ the references for the optical classes are as follows: We used the PW \citep{1981MNRAS.194..331P} optical identifications where available, otherwise  the PR \citep{1981ApJ...248...61P} and CJ1 \citep{1995ApJS...98....1P}, where we replaced  ``SO'' and  ``BL'' entires with ``Q''. The only exceptions are $^a$ from \citet{1994AAS..105..211S}, and $^b$ from \citet{2000ApJ...534...90P}. The  columns are as follows: Source Names (1--3), Membership in the PR, CJ1 and PW Samples (4--6), Source Type (7), The Types listed in column (7)  are as follows: Quasars (Q), galaxies (G), bona fide CSO (CSO), Fanaroff and Riley types~I, II and intermediate (FR~I, FR~II, FR?); objects unresolved by \citet{1981MNRAS.194..331P} on the 5\,km~Telescope (U,CSS or U,CFS for compact steep spectrum and compact flat spectrum sources, respectively), doubles indentified by \citet{1981MNRAS.194..331P} in which the optical ID coincides with one of the radio components (D2); Compact Steep Spectrum objects identified by \citet{1981MNRAS.194..331P} (CSS); Objects found to be compact in various VLBI surveys  other than any identified by \citet{1981MNRAS.194..331P} (CT,CSS or CT,CFS) for compact steep spectrum and compact flat spectrum sources, respectively;  CSO Catalog ID (8), Optical Identifications (9), and Structure References (10+11). No attempt has been made at complete map references for each object since these number in the tens for many objects and the purpose of this Table is solely to provide justification for the claim that the structures of all of these sources are well known.  References: 1 \citet{1968MNRAS.138..259M}; 2 \citet{1972MNRAS.156..377B}; 3 \citet{1974MNRAS.169..477P}; 4 \citet{1975MmRAS..80..105R}; 5 \citet{1975MNRAS.173..309L}; 6 \citet{1975AandA....38..381M};7 \citet{1976Natur.262..179B}; 8 \citet{1977MmRAS..84...61J};  9 \citet{1981MNRAS.194..331P}; 10 \citet{1977MNRAS.178..577C}; 11 \citet{1977MNRAS.181..465W}; 12 \citet{1980AJ.....85..981F};13 \citet{1980Natur.287..208B}; 14 \citet{1981ApJ...248...61P}; 15 \citet{1981Natur.294...47P}; 16 \citet{1981AandAS...43..381K}; 17 \citet{1981MNRAS.195..261L}; 18 \citet{1981MNRAS.197..253B}; 19 \citet{1982MNRAS.198..843P}; 20 \citet{1983MNRAS.204..151L}; 21 \citet{1984AandA...135..289R}; 22 \citet{1984AJ.....89..932O};
23 \citet{1984AJ.....89.1478S}; 24 \citet{1980AandA....85...36S}; 25 \citet{1984MNRAS.208..545L}; 26 \citet{1982MNRAS.200..705J}; 27 \citet{1984ApJ...282L..55V}; 28  \citet{1986AandAS...65..145F}; 29 \citet{1986MNRAS.222..753L}; 30 \citet{1987MNRAS.225....1A}; 31 \citet{1988ApJ...328..114P}; 32 \citet{1989AJ.....97..674B}; 33 \citet{1990MNRAS.246..477P}; 34 \citet{1991AJ....102..537L}; 35 \citet{1995AandAS..114..197A}; 36 \citet{1975MNRAS.172..181T}; 37 \citet{1994AJ....108..766B}; 38 \citet{1994AJ....108..821L}; 39 \citet{1994ApJ...435..116G}; 40  \citet{1994AandAS..105..247A}; 41  \citet{1995ApJS...98....1P} or \citep{1995ApJS...98...33T}; 42  \citet{1995ApJS...99..297X}; 43 \citet{1995AandAS..112..235A}; 44 \citet{1995AandAS..110..213R}; 45 \citet{1995AandA...295..629S}; 46 
 \citet{1996MNRAS.278..273H}; 47  \citet{1997MNRAS.291...20L}; 48 \citet{1998MNRAS.299..467L}; 49 \citet{1995ApJS...99..349N}; 50 \citet{1999MNRAS.304..271D}; 51 \citet{2000ApJS..131...95F}; 52 \citet{2000ApJ...534..172L}; 53 \citet{2001MNRAS.321...37S}; 54 \citet{2013MNRAS.433..147D}:55 \citet{2006MNRAS.372.1607T};
  56 \citet{1995AA...295...27D}; 57 \citet{2021MNRAS.504.2312D}; 58 MOJAVE
website: \url{https://www.cv.nrao.edu/MOJAVE/allsources.html}; 59 Radio Fundamental Catalog (RFC); 60 \citet{1998AJ....115.1295K}; 61 \citet{2018NatAs...2..472G}; 62
\citet{2014ApJS..212...19F}; 63 \citet{1991MNRAS.250..215A}; 64 \citep{1997MNRAS.288..859H};65 \citet{2016MNRAS.459..820T}; 66
\citet{1990AA...232...19B}; 67
\citet{2011AA...535A..24S}; 68
\citet{1995AAS..114..197A}; 69
\citet{1995PhDT.........2X}; 70
\citet{1995AA...295...27D}; 71
\citet{2005AA...443..891S}; 72
\citet{2002AA...385..768X}; 73
\citet{2006ApJ...648..148N}; 74
\citet{2014MNRAS.438..463O}; 75
\citet{1996ApJ...460..612R}
}
\end{deluxetable*}

\clearpage
\bibliographystyle{aasjournal}





\end{document}